\documentclass[epj]{svjour}
\usepackage{graphics}
\usepackage{graphicx}
\usepackage{amssymb}
\usepackage{amsmath}
\usepackage{xspace} 
\usepackage{color}
\renewcommand{\vec}[1]{\mathbf{#1}}
\begin{document}
\title{Linear response within the projection-based renormalization method}
\subtitle{Many-body corrections beyond the random phase approximation}
\author{Van-Nham Phan\inst{1} \and Holger Fehske\inst{2} \and Klaus W. Becker\inst{3}}
\institute{Institute of Research and Development, Duy Tan University, K7/25 Quang Trung, Danang, Vietnam \and
Institut f{\"u}r Physik, Ernst-Moritz-Arndt-Universit{\"a}t Greifswald, D-17489 Greifswald, Germany \and
Institut f{\"u}r Theoretische Physik, Technische Universit{\"a}t Dresden, D-01062 Dresden, Germany
}
\date{Received: date / Revised version: date}
%

\abstract{
The explicit evaluation of linear response coefficients for interacting many-particle systems still poses a considerable challenge to
theoreticians.  In this work we use a novel many-particle renormalization technique, the so-called projector-based renormalization method, to
show how such coefficients can systematically be evaluated. To demonstrate the prospects and power of our approach we consider 
the dynamical wave-vector dependent spin susceptibility of the two-dimensional Hubbard model and 
also determine the subsequent magnetic phase diagram close to half-filling.   We show that the superior treatment
of (Coulomb) correlation and fluctuation effects within the projector-based renormalization method  significantly improves the standard random phase approximation results. 
%
} 
\maketitle
\section{Introduction}
\label{I}

The most popular approach to evaluate linear response coefficients for many-body systems 
is probably the standard random phase approximation (RPA), which is used in quite different fields in physics.  
Besides its physical merits 
including the fulfillment of conservation laws  the popularity of the RPA results from is conceptual  
simplicity in its derivation as well as from its numerical  practicability. 
Attempts to go beyond the RPA have turned out to be extremely demanding.  
One of the options to improve the RPA are methods  
based on conserving approximations (see, e.g.,~Ref.~\cite{BSW89}), thereby following an approach which was 
introduced by Baym and Kadanoff~\cite{BK61,KB62}.   
Conserving approximations are consistent with microscopic conservation laws for particle 
number, energy or momentum. Other work is based on the  
time-dependent Hartree-Fock approximation which uses a frequency-dependent local field factor in a 
modified RPA expression~\cite{GV08}. However, this method turns out to be rather complex 
and physically unsatisfactory. Attempts to  
find a numerical solution of the basic integral equation could not be reached 
without further approximations
(see discussion in Ref.~\cite{GV08} and references therein). 
Recently, Vilk and Tremblay extended the RPA by including vertex corrections, 
taken into account correlation and exchange effects~\cite{VT97}. Comparing their {\it ansatz} for  
double occupancies in the Hubbard model, the authors found good quantitative agreement with 
results from Monte Carlo simulations for single-particle and two-particle properties~\cite{BSW94,VCT94}. 
Another way to improve the RPA is the so-called self-consistent RPA~\cite{JSDB05}, which is based on 
a non-perturbative variational scheme.  This approach has been adopted to the investigation of 
various nontrivial models but is however limited to small systems~\cite{JSDB05,KS94,HMDS02}.

For these reasons it is important to develop new many-particle techniques having the ability to  
include correlation effects. One approach that overcomes some of the shortcomings
of the RPA is the projection-based renormalization method (PRM)~\cite{BHS02,SHB09b}.  In the recent past the PRM 
has been successfully  applied to several physical problems such as 
superconductivity~\cite{HB03}, quantum phase transitions in coupled electron-phonon systems \cite{SHBWF05,SHB06,PFB11,PBF13}, exciton and plasmaron formation~\cite{PFB11,PF12}, 
BCS-BEC transition~\cite{PBF10}, electronic phase separation~\cite{ESBF12}, valence transitions~\cite{PMB10}, or the Kondo lattice problem~\cite{SB13}.
In the present work, adding time- and wave-vector-dependent  external fields, 
we demonstrate how the PRM can be combined with linear response theory
in order to calculate response functions for generic correlation models. In particular we derive an explicit analytical expression for the dynamical spin susceptibility
of the Hubbard model. For the two-dimensional (2D) case the PRM phase boundaries between the paramagnetic and 
antiferromagnetic respectively ferromagnetic phases are determined for weak-to-intermediate Hubbard interactions.
An elaborate weak-coupling approach is of particular importance in low spatial dimensions since in 1D and 2D  also 
weakly interacting systems tend to be strongly correlated. 

The paper is organized as follows. In the Sect.~\ref{II}, we recapitulate the RPA  to the Hubbard model. 
Section~\ref{III} introduces the PRM approach, which is applied to the Hubbard model in Sect.~\ref{IV},
focusing on the response to an external magnetic field. Thereby  the 
renormalization equations for the model parameters, the transformations of the operators  and various expectations values 
are derived.  Details can be found in Appendices~A and~B. Section~\ref{IV.4} provides our main analytical result: the explicit expression for the 
dynamical spin susceptibility. Selected numerical results for the 2D Hubbard model
can be found in Sec.~\ref{V},  in particular the ground-state phase diagram in the $U$-$n$ plane and the wave-vector- and frequency-dependence
of the magnetic susceptibility. We conclude in Sect.~\ref{VI}.
 
\section{Standard RPA approach to the Hubbard model}
\label{II}
The Hubbard model is a paradigmatic model for the study  of correlation effects in itinerant electron systems. 
Independently proposed by Gutzwiller~\cite{Gu63}, Hubbard~\cite{Hu63}, and Kanamori~\cite{Ka63} in 1963, it was originally 
designed to describe the ferromagnetism of transition metals. Successively, the model has been studied 
in the context of antiferromagnetism, metal-insulator transition, and high temperature superconductivity.
The Hubbard Hamiltonian is given by
\begin{eqnarray}
\label{1}
\mathcal H &=&  \bar{t} \sum_{\langle i, j \rangle\sigma} c_{i\sigma}^\dag c^{}_{j\sigma} 
+ \frac{U}{2} \sum_{i \sigma} n^{}_{i\sigma} n^{}_{i, -\sigma}\,.
\end{eqnarray}
Here $c_{i\sigma}^\dag$ $(c_{i\sigma}^{})$ is a fermionic creation (annihilation) operator of a spin 
$\sigma\;(=\uparrow,\downarrow)$ electron, and $n^{}_{i\sigma} = c_{i\sigma}^\dag c_{i\sigma}^{}$.
$U$ denotes the on-site Coulomb interaction and $\bar{t}$ are the electron transfer matrix elements
between nearest-neighbor Wannier sites $i$ and $j$.   The physics of the model is governed by the competition 
between itinerancy ($\bar t$; delocalization, kinetic energy) and short-range Coulomb repulsion ($U$; localization, magnetic order),
where the fermionic nature of the charge carriers is of great importance (Pauli exclusion principle). Besides the parameter ratio $U/\bar{t}$,
the particle density $n$, the temperature $T$, and the spatial dimension $D$ (geometry of the lattice) are crucial. 

Although a tremendous amount of work has been devoted to the Hubbard model, in order to determine its ground-state, spectral and thermodynamic 
properties, exact results are rare and only a few special cases and limits are ultimately understood.
In 1D, the  algebraic and thermodynamic Bethe {\it ansatz} enables an exact treatment of the model~\cite{Tak99,EFGKK05}. 
However the Bethe {\it ansatz} technique does not provide a complete framework since it generally does not allow the evaluation of the response functions. 
For $D>1$  approximations are unavoidable anyway. There usually the weak- $(U/W \ll 1)$  and strong-coupling $(U/W \gg 1$) limits of the model were studied, 
with uncertain extrapolations to the region $U/W \sim 1$.  Here $W$ is the bare electronic bandwidth. For a $D$-dimensional hypercubic lattice we
have $W=4D\bar{t}$. 

In consideration of  the magnetic behavior of a Hubbard model system 
the response to an applied external field is of particular importance. 
Adding a small  magnetic field that periodically oscillates in space and time  
the Hamiltonian takes the form 
\begin{eqnarray}
\label{3}
\mathcal H(t) = \mathcal H_{kin}  + \mathcal H_U + {\mathcal H}_h (t)\,,
\end{eqnarray}
where 
\begin{eqnarray}
\label{4}
\mathcal H_{kin} &=&  \sum_{\mathbf  k \sigma} {\varepsilon}_{\vec k} \, c_{\vec k\sigma}^\dag c^{}_{\vec k\sigma}, 
\end{eqnarray}
is the kinetic energy of electrons in momentum space. In the numerical 
evaluation below the Hubbard model is considered on a square, where the dispersion is given by 
\begin{equation}\label{4a}
{\varepsilon}_{\vec k}=2\bar{t}(\cos k_x+\cos k_y) -\mu 
\end{equation} 
with chemical potential $\mu$. The last two terms in Eq.~(\ref{3}) respectively read
\begin{eqnarray}
\mathcal H_U& = &\frac{U}{2N} \sum_{\vec k \vec k' \vec p \sigma} c_{\vec k \sigma}^\dag\, c^{}_{\vec k + \vec p, \sigma} \,
c_{\vec k', -\sigma}^\dag \, c^{}_{\vec k' -\vec p, -\sigma} \, , 
\end{eqnarray}
and
\begin{eqnarray}
\label{5}
{\mathcal H}_h(t) &=&
 - \sum_i  h(t) \cos{(\vec q \cdot  \vec R_i)} \, s^z_i  =  - \frac{ h(t)}{2}\big(s^z_{\vec q} + s^z_{-\vec q} \big) 
 \nonumber \, . \\
&&
 \end{eqnarray}
 Note that the wave vector $\vec q$ is imposed by the external field.   
 $s_{\vec q}^z$ is the component of the spin operator in field direction:
 \begin{equation}
s^z_{\vec q} = \sum_i e^{i \vec q \cdot \vec R_i} s_i^z=  \sum_{\vec k \sigma} \frac{\sigma}{2} c_{\vec k \sigma}^\dag c^{}_{\vec k - \vec q,  \sigma} \, .
\label{6}
\end{equation} 
 Then the linear response of the spin expectation value $\langle s^z_{-\vec q} \rangle(t)$ with respect to $h(t) \sim \textrm{Re} \;e^{-i\omega t}$ is given by
 \begin{eqnarray}
 \label{7}
\langle s^z_{-q} \rangle(t) &=& -{i}  \int_0^\infty dt'  \langle [s_{-\vec q}^z(t') , {\mathcal H}_h(t- t')] \rangle
\nonumber \\
&=&  
 N \chi(\vec q, \omega) \frac{h(t)}{2} \, , 
\end{eqnarray} 
where
\begin{eqnarray}
 \label{8}
\chi(\vec q, \omega) &=& \frac{i}{N}\int_0^\infty dt'\langle [s^z_{-\vec q}(t') , s^z_{\vec q} ] \rangle \, 
e^{i (\omega + i \eta) t'} 
\end{eqnarray} 
($\eta = 0^+$) is the formal expression for the dynamical magnetic susceptibility. 
Here, the expectation value is formed  
with Hamiltonian $\mathcal H$ [Eq.~(\ref{1})] in the absence of the external perturbation. Due to the 
Coulomb part $\mathcal H_U$
in $\mathcal H$ a straightforward evaluation of $\chi(\vec q, \omega)$ turns out to be difficult.
To proceed, in a first step, let us  introduce fluctuation operators 
\begin{eqnarray}
\label{9}
: c_{\vec k \sigma}^\dag\, c^{}_{\vec k + \vec p, \sigma}: &=& 
c_{\vec k \sigma}^\dag\, c^{}_{\vec k + \vec p, \sigma} - 
\langle c_{\vec k \sigma}^\dag\, c^{}_{\vec k + \vec p, \sigma} \rangle(t) \, , \\
&& \nonumber
 \end{eqnarray}
where the expectation value $\langle \cdots \rangle(t)$ on the right-hand side is formed with 
$\mathcal H(t)$  and therefore becomes time dependent. It can be simplified 
by help of  the operator identity 
 \begin{eqnarray*}
  c_{\vec k \sigma}^\dag\, c^{}_{\vec k + \vec p,  \sigma}  &=&
\frac{1}{2} \sum_{\tilde{\sigma}} c^\dag_{\vec k \tilde \sigma} c^{}_{\vec k + \vec p, \tilde{\sigma}} 
+ \sigma \sum_{\tilde \sigma }\frac{\tilde \sigma}{2} \, 
 c_{\vec k \tilde{\sigma} }^\dag c^{}_{\vec k + \vec p, \tilde \sigma} 
\nonumber \\
&&
 \end{eqnarray*}
to give
  \begin{eqnarray}
  \label{10}
 \langle c_{\vec k \sigma}^\dag\, c^{}_{\vec k + \vec p,  \sigma}\rangle(t)  &=&
\frac{\delta_{\vec p,0}}{2} \sum_{\tilde{\sigma}} 
\langle c^\dag_{\vec k \tilde \sigma} c^{}_{\vec k , \tilde{\sigma}} \rangle 
+\sigma \delta_{\vec p, - \vec q} \langle  s^z_{\vec k, \vec q}\rangle(t) \nonumber \\
&+&  \sigma \delta_{\vec p,  \vec q}  \langle  s^z_{\vec k, -\vec q}\rangle(t) 
   \, .
  \end{eqnarray}
Here $\vec p$ is either $\vec p =\pm \vec q$ or $\vec p = 0$. 
We have also introduced the $\vec k$-resolved spin operator 
\begin{equation}
\label{11}
s^z_{\vec k,\vec q} = \sum_{\tilde \sigma} \frac{\tilde \sigma}{2} c^\dag_{\vec k \tilde \sigma}c^{}_{\vec k - \vec q, {\tilde \sigma}} \, .
\end{equation}
The expectation value $\langle c^\dag_{\vec k\sigma} c^{}_{\vec k \sigma} \rangle$ 
has no linear contribution in $ h(t)$ and can be considered as time independent.  

Using Eqs.~(\ref{9}) and (\ref{10}), the Hamiltonian $\mathcal H(t)$ can be rewritten as 
\begin{eqnarray}
\label{12}
\mathcal H(t) = {\mathcal H}_{0} +  \hat {\mathcal H}_h(t) + \mathcal H_{f}(t)\,,
\end{eqnarray}
where 
\begin{eqnarray}
 \label{13}
 && {\mathcal H}_{0} = \sum_{\vec k \sigma} \Big( \varepsilon^{}_{\vec k} + \frac{U}{2} \langle n \rangle
\Big) \, c_{\vec k \sigma}^\dag c^{}_{\vec k \sigma}\,, \\
& & \hat {\mathcal H}_h(t) 
= -  \frac{\hat  h_{\vec q}(t)}{2} \,  \Big( s^z_{\vec q} + s^z_{-\vec q}\Big)  \nonumber  \,,
\end{eqnarray}
and
\begin{eqnarray}
\label{14}
\frac{{\hat h}_{\vec q}(t)}{2} =    \frac{ h(t)}{2} + \frac{U}{N} \langle s^z_{ -\vec q} \rangle(t)  \, .
\end{eqnarray}
In Eq.~(\ref{14}) we have introduced an effective field ${\hat h}_{\vec q}(t)$ that contains  an internal field  proportional to $U$. Also the kinetic energy  ${\mathcal H}_{0}$ 
has acquired a Hartree shift proportional to $U$.  Finally, the part $\mathcal H_f(t)$ reads
 \begin{eqnarray}
\label{15}
\mathcal H_{f}(t) = \frac{U}{2N} \sum_{\vec k \vec k' \vec p \sigma} 
: c_{\vec k \sigma}^\dag\, c^{}_{\vec k + \vec p, \sigma}: \,
:c_{\vec k', -\sigma}^\dag \, c^{}_{\vec k' -\vec p, -\sigma} :  \, ,\quad 
\end{eqnarray}
where the $t$-dependence enters via the fluctuation operators. 

 The standard RPA expression for $\chi(\vec q, \omega)$ 
 is obtained by neglecting the fluctuating part $\mathcal H_{f}(t)$ completely, i.e., $\mathcal H(t)$
 reduces to
 \begin{eqnarray}
\label{16}
\mathcal H_{RPA}(t) &=&  {\mathcal H}_{0} +  \hat {\mathcal H}_h(t) \, .
 \end{eqnarray}
The linear response of $ \langle s^z_{-q} \rangle(t)$ to the effective field $\hat h_{\vec q}(t)$ becomes  
 \begin{eqnarray}
 \label{17}
 \langle s^z_{-q} \rangle(t) &=& -{i}  \int_0^\infty dt'  \langle [s_{-\vec q}^z(t') , \hat{\mathcal H}_h(t- t')] \rangle_{0}
 \nonumber \\
 &=&  N \chi^{}_{0} (\vec q, \omega) \,   \frac{\hat h_{\vec q}(t)}{2}  \, ,
 \end{eqnarray}
 where the expectation value is now formed with the unperturbed Hamiltonian $\mathcal H_{0}$.
 By help of  relation (\ref{14}) one arrives at 
   \begin{eqnarray}
 \label{18}
 \langle {s}^z_{-\vec q}\rangle (t) &=&  \frac{N \chi^{}_{0} (\vec q, \omega)}{1- 
 \displaystyle U \chi^{}_{0} (\vec q, \omega) } \,  \frac{ h(t)}{2}  
 \,  .
 \end{eqnarray}
Here $\chi_{0} (\vec q, \omega)$ is the dynamical susceptibility of the unperturbed system 
${\mathcal H}_{0}$
 \begin{eqnarray}
 \label{19}
 \chi^{}_{0} (\vec q, \omega) &=& \frac{i}{N} \int_0^\infty dt'
 \langle [s^z_{-\vec q}(t'), s^z_{\vec q}  ] \rangle_{0} \, e^{i (\omega + i\eta)t'}   \nonumber \\
&=& \frac{1}{2N} \sum_{\vec k} \frac{f(\varepsilon_{\vec k + \vec q}) -f(\varepsilon_{\vec k})}{\varepsilon_{\vec k}- \varepsilon_{\vec k + \vec q} + \omega + i \eta} 
 \end{eqnarray}
with $\eta =0^+$.  Note that Eqs.~(\ref{17})-(\ref{19}) are the usual RPA equations.Thus, the dynamical RPA susceptibility  is defined 
by the prefactor in  Eq.~(\ref{18}):
\begin{equation}
 \label{20}
 \chi^{}_{RPA} (\vec q, \omega) =   \frac{ \chi^{}_{0} (\vec q, \omega)}{1- 
 \displaystyle {U} \chi^{}_{0} (\vec q, \omega) } \, .
\end{equation}
 
\section{PRM formalism}
\label{III}

Our aim is to evaluate the dynamical susceptibility $\chi(\vec q, \omega)$ beyond the standard 
RPA  by including fluctuation processes, which are induced by  the fluctuation part $\mathcal H_f(t)$
of the Coulomb interaction.   To this end, we combine  linear response theory with the PRM. 
The PRM in the original version without time-dependent external field starts 
by separating a given many-particle Hamiltonian into an unperturbed part $\mathcal H_0$ 
and a time-independent perturbation $\mathcal H_f$. 
Since $\mathcal H_f$ and $\mathcal H_0$ do not commute, the perturbation induces transitions between the eigenstates of 
$\mathcal H_0$. The basic idea of the PRM is to eliminate successively all transitions due to $\mathcal H_f$ so that finally only the 
unperturbed, yet renormalized  Hamiltonian (now called $\tilde{\mathcal H}_0$) remains.
 
In the present case the many-particle Hamiltonian (\ref{12}) is time dependent,
 $ {\mathcal H}(t) = \mathcal H_{0} + \hat{\mathcal H}_{h}(t) + \mathcal H_f(t)$, since $\hat{\mathcal H}_{h}(t)$
 is time-dependent due to the external field. As before, our aim is to evaluate the response of the expectation 
 value $\langle s_{-\mathbf q}\rangle(t)$ up to linear order in the external field. However, the fluctuation  term $\mathcal H_f(t)$ 
of the Coulomb interaction should now be taken into account. 
 Since both  $ \mathcal H_{0}$ and  $\hat{\mathcal H}_{h}(t)$ do not commute with 
  $\mathcal H_f(t)$, the latter Hamiltonian will  henceforth be  considered as perturbation. In particular, $\mathcal H_f(t)$ again
 induces transitions between the eigenstates of $\mathcal H_{0}$.  
 In the PRM these transitions will be eliminated by 
 a sequence of unitary transformations, which are performed in small steps $\Delta \lambda$ 
 by proceeding from large to small transition energies. Let $\mathcal H_\lambda$ be the Hamiltonian 
 after all transitions with energies larger than some cutoff $\lambda$ have already been integrated out. 
 The transformation from cutoff $\lambda$ to a somewhat reduced cutoff $\lambda -\Delta \lambda$ formally reads
 \begin{equation}
 \label{21}
 {\mathcal H}_{\lambda - \Delta \lambda}(t) = e^{X_{\lambda, \Delta \lambda}} {\mathcal H}_\lambda(t) \,
 e^{-X_{\lambda, \Delta \lambda}} \, .
 \end{equation}
 Here $X_{\lambda, \Delta \lambda}= - X^\dag_{\lambda, \Delta \lambda}$ 
 is the generator of the unitary transformation from $\lambda$ to $\lambda - \Delta \lambda$, 
whereas $\mathcal H_\lambda(t)$,
 \begin{equation}
 \label{22}
 {\mathcal H}_{\lambda}(t) = \mathcal H_{0,\lambda} + \hat{\mathcal H}_{h,\lambda}(t) +
 \mathcal H_{f,\lambda}(t) \, ,   
  \end{equation}
  represents the renormalized Hamiltonian
 after all transitions (in the 
 eigenbasis of ${\mathcal H}_{0,\lambda}$)  with energies larger than 
  $\lambda$ have been eliminated  from ${\mathcal H}_f(t)$. Similarly, 
 ${\mathcal H}_{\lambda - \Delta \lambda}(t)$ denotes  the Hamiltonian 
 with the somewhat reduced cutoff $\lambda - \Delta \lambda$. 
 Due to transformation (\ref{21})  the parameters of $\mathcal H_\lambda$
 become renormalized, and also  new terms can in principle be generated. 
 For the generator $X_{\lambda, \Delta \lambda}$ we chose 
  \begin{eqnarray}
 \label{23}
 X_{\lambda, \Delta \lambda}(t) &=& \frac{1}{\mathbf L_{0,\lambda}} \mathbf Q_{\lambda -\Delta \lambda} 
\,  \mathcal H_{f,\lambda}(t) \, ,
  \end{eqnarray}
 which agrees with the lowest order result for $X_{\lambda, \Delta \lambda}$ in the absence of 
 an external field~\cite{BHS02}. Here $\mathbf L_{0,\lambda}$ is the Liouville operator of the 'unperturbed' Hamiltonian
  $\mathcal H_{0,\lambda}$. It is defined by the commutator $\mathbf L_{0,\lambda}\, {\mathcal A} =
  [\mathcal H_{0,\lambda}, {\mathcal A}]$, applied to any operator variable ${\mathcal A}$. Moreover, 
  $\mathbf Q_{\lambda -\Delta \lambda}$  is a generalized projection operator. It projects 
  on all transition operators in $\mathcal H_{f,\lambda}$  (in the basis of $\mathcal H_{0,\lambda}$) 
  with transition energies larger than  $\lambda -\Delta \lambda$. Note that $X_{\lambda, \Delta \lambda}$ also
  depends on time $t$ since $\mathcal H_{f,\lambda}(t)$ depends on $t$ via the external field. 
  
The elimination procedure starts from the original Ha\-miltonain
$\mathcal H(t)$ (where the largest cutoff energy is called  $\lambda= \Lambda$) 
and proceeds in steps $\Delta \lambda$ until 
$\lambda=0$ is reached. This limit   provides the 
desired effective Hamiltonian \\
$\tilde{\mathcal H}(t):=\mathcal H_{\lambda \rightarrow 0}(t)$ with 
${\mathcal H}_{\lambda \rightarrow 0}(t) = \mathcal H_{0, \lambda\rightarrow 0}+ 
\mathcal H_{h,\lambda \rightarrow 0}(t)$. Note that  
the elimination of the transitions leads to renormalized parameters in
$\tilde{\mathcal H}(t)$.
Thus, after all transitions from ${\mathcal H}_f(t)$ have been used up, the final Hamiltonian 
 $\tilde{\mathcal H}(t)$ is diagonal or at least quasi-diagonal and allows to evaluate expectation values. 
As a matter of course the parameters  in  $\tilde{\mathcal H}(t)$ depend on their
values in the original model $\mathcal H(t)$. 

Having in mind small renormalization steps $\Delta \lambda$, the transformation (\ref{21}) 
from $\lambda$ to $\lambda - \Delta \lambda$ can be restricted 
to an expansion up to second order in $U$ (and linear order in $h(t)$). 
Then $\mathcal H_{\lambda-\Delta \lambda}(t)$ reads 
\begin{eqnarray}
 \label{24}
&& {\mathcal H}_{\lambda - \Delta \lambda} = e^{X_{\lambda, \Delta \lambda}} \mathcal H_\lambda 
e^{-X_{\lambda, \Delta \lambda}} \nonumber \\
&& \quad = \mathcal H_{0,\lambda} + 
\hat{\mathcal H}_{h,\lambda} +
  \mathbf P_{\lambda -\Delta \lambda} \mathcal H_{f,\lambda}  
  +  [X_{\lambda, \Delta \lambda}, \hat {\mathcal H}_{h, \lambda} ]   \nonumber \\
 && \quad + 
   [X_{\lambda, \Delta \lambda}, {\mathcal H}_{f,\lambda}]   - 
    \frac{1}{2}     [X_{\lambda, \Delta \lambda},  
    \mathbf Q_{\lambda -\Delta \lambda}  {\mathcal H}_{f,\lambda}]
    \nonumber \\
  && \quad  + \frac{1}{2} [ X_{\lambda, \Delta \lambda}, 
  [ X_{\lambda, \Delta \lambda},  \hat {\mathcal H}_{h,\lambda}]  ] 
  + \cdots \, ,
  \end{eqnarray} 
where relation (\ref{23}) was used and the explicit $t$-dependence is suppressed. 
$\mathbf P_{\lambda -\Delta \lambda}= 
1-  \mathbf Q_{\lambda -\Delta \lambda}$ 
is the projector  on all low-energy transitions
with energies smaller than $\lambda- \Delta \lambda$. 
The  commutators in Eq.~(\ref{24}) give rise to renormalization 
contributions to $\mathcal H_{\lambda- \Delta \lambda}(t)$. Having in mind an application of linear response theory, 
Eq.~\eqref{24} has to be evaluated up to linear order in the external field. Finally, comparing the result of 
the evaluated right-hand side with the generic form of $\mathcal H_\lambda$ one is led to renormalization equations, 
which relate the parameters of the Hamiltonian at cutoff $\lambda$ with those at cutoff $\lambda - \Delta \lambda$. 
  
As is discussed below one also has to  evaluate expectation values, which are 
 formed with Hamiltonian $\mathcal H$. Exploiting the unitary invariance of operators below a trace 
we can write
\begin{eqnarray}
\label{25}
 \langle {\mathcal A} \rangle =
\frac{{\mbox{Tr}} {\mathcal A} e^{-\beta {\mathcal H}}}{ {\mbox{Tr}} e^{-\beta {\mathcal H}}} = 
\langle {\mathcal A}(\lambda) \rangle_{{\mathcal H}_\lambda}=
 \langle \tilde{{\mathcal A}}\rangle_{\tilde{{\mathcal H}}} \, , 
\end{eqnarray}
where the same unitary transformation as before is applied to operator $\mathcal A$, 
i.e.~${\mathcal A}(\lambda) = e^{{X}_\lambda}{\mathcal A}e^{-{X}_\lambda}$. 
Here ${X}_\lambda$ is generator of the unitary 
transformation between cutoff $\Lambda$ and $\lambda$, and $\tilde{{\mathcal A}}= 
{\mathcal A}(\lambda \rightarrow 0)$. 
Thus, additional renormalization equations for  ${\mathcal A}(\lambda)$
are required. \\

Let us mention that  Wegner and coworkers~\cite{We94,Keh06} have introduced a theoretical approach 
related to the PRM.  This approach is based on the application of continuous unitary transformations instead of discrete ones 
as in the present case. To our knowledge it was not applied up to now to the investigation of 
many-body corrections beyond the random phase approximation  discussed in the present work.
However,  correlation and fluctuation processes can be discussed in this method as well, compare for instance references  \cite{ZaDo11,FrKe10,KDU12}, or~\cite{VMM13}.  The  relationship between the continuous method and the PRM for the case without time-dependent field is studied
in references  \cite{PBF10}  and \cite{HSB08}. 
There it was shown that the continuous method can be derived within the PRM framework in the limit of small $\Delta \lambda \rightarrow 0$ 
using a particular choice for the complement part ${\mathbf P}_{\lambda- \Delta \lambda}X_{\lambda, \Delta \lambda}$ of generator (\ref{23}).

\section{PRM for the Hubbard model}
\label{IV}

\subsection{{\it \bf Ansatz} for  Hamiltonian $H_\lambda(t)$}
\label{IV.1}

We are now in the position to apply the general formalism of Section \ref{III} to the 
Hubbard model. Thereby,  the influence of the fluctuation term $\mathcal H_f (t)$ will be 
investigated. Following the ideas of the PRM, we have to start from an {\it ansatz} for the renormalized 
Hamiltonian $\mathcal H_\lambda(t)$. A perturbative evaluation of transformation (\ref{21})
suggests the use of the following expression for ${\mathcal H}_{\lambda}(t)$ (see Appendix A),
where 
\begin{eqnarray}
\label{26}
{\mathcal H}_{0,\lambda} &=&  \sum_{\vec k \sigma} {\varepsilon}^{}_{\vec k,\lambda} \, 
c_{\vec k\sigma}^\dag  c^{}_{\vec k \sigma}  \,,
\label{27}\\
\hat {\mathcal H}_{h, \lambda}(t) &=& 
-\sum_{\vec k \sigma} \Big[ \Big( \frac{\hat h_{\vec q}(t)}{2} + u_{\vec k,\vec k -\vec q,\lambda}(t) \Big)\, 
\frac{\sigma}{2} c_{\vec k \sigma}^\dag c^{}_{\vec k - \vec q, \sigma} 
\label{27a}\\ 
&&\qquad\qquad+ \textrm{H.c.}\Big] \nonumber
 \\
 && -
   \frac{1}{N} \sum_{\vec k \vec k' \vec p \sigma}  \Big[
   v_{\vec k, \vec k+\vec p -\vec q; \vec k', \vec k'- \vec p , \lambda}(t) \frac{\sigma}{2}
 \nonumber 
  \\
&& \times   
    :c_{\vec k \sigma}^\dag c^{}_{\vec k + \vec p -\vec q, \sigma}: \, 
   :c_{\vec k', -\sigma}^\dag c^{}_{\vec k'- \vec p, -\sigma} :  +  \textrm{H.c.} \Big] \,,\nonumber \\
 \label{28}
  {\mathcal H}_{f, \lambda}(t) &=& 
 \frac{U}{2N} \sum_{\vec k \vec k' \vec p \sigma}
\mathbf P_{\lambda} \big( : c^\dag_{\vec k \sigma} \,c^{}_{\vec k + \vec p, \sigma}: \, :c^\dag_{\vec k', -\sigma} \,
c^{}_{\vec k' - \vec p, -\sigma} : \big) \, .
\nonumber\\&& 
 \end{eqnarray} 
 Due to the renormalization all coefficients in $ {\mathcal H}_{0,\lambda}$ and 
$\hat {\mathcal H}_{h, \lambda}(t)$  depend on $\lambda$.
Moreover, in $\hat {\mathcal H}_{h,\lambda}(t)$ new operator contributions 
are generated.

The coefficients $u_{\vec k,\vec k - \vec q,\lambda}(t)$ and 
   $ v_{\vec k, \vec k+\vec p -\vec q; \vec k', \vec k'- \vec p , \lambda}(t)$ 
  are expected to  depend linearly on the external field and are therefore
   explicitly time-dependent. From hermiticity follows that they obey the relations:
   \begin{eqnarray}
u_{\vec k, \vec k-\vec q,\lambda}&=& u^{*}_{\vec k +\vec q, \vec k,\lambda},\\
v_{\vec k,\vec k +\vec p -\vec q; \vec k', \vec k' -\vec p, \lambda}&=&
 - v^*_{\vec k'- \vec p, \vec k'; \vec k +\vec p +\vec q,\vec k,\lambda}\nonumber\\ &=&
 - v_{\vec k', \vec k' -\vec p; \vec k,\vec k +\vec p -\vec q, \lambda}\,.
\end{eqnarray}

Finally  $\mathbf P_{\lambda}$ in $\mathcal H_{f,\lambda}(t)$  projects on the 
low-energy excitations smaller than $\lambda$. 
One finds
 \begin{eqnarray}
 \label{29}
 && {\mathcal H}_{f, \lambda}(t) = \frac{U}{2N} \sum_{\vec k \vec k' \vec p \sigma}  \Big( \Theta_{\vec k, \vec k+\vec p; \vec k', \vec k' -\vec p, \lambda } \,\\
&& \quad \times 
c^\dag_{\vec k \sigma} \,c^{}_{\vec k + \vec p, \sigma}  \, c^\dag_{\vec k', -\sigma} \,
c^{}_{\vec k' - \vec p, -\sigma} \nonumber \\
&& \quad - \Theta_{\vec k, \vec k + \vec p,\lambda} \, 
c^\dag_{\vec k \sigma} \,c^{}_{\vec k + \vec p, \sigma}  \, \langle c^\dag_{\vec k', -\sigma} \,
c^{}_{\vec k' - \vec p, -\sigma}\rangle  \nonumber \\
 &&\quad - \Theta_{\vec k', \vec k' - \vec p,\lambda} \,
 \langle c^\dag_{\vec k \sigma} \,c^{}_{\vec k + \vec p, \sigma} \rangle  \,  
 c^\dag_{\vec k', -\sigma} 
 c^{}_{\vec k' - \vec p, -\sigma}   
 + \textrm{const.} \Big) \nonumber \, ,
 \end{eqnarray} 
 which shows that the operators on the right-hand side have different 
 transition energies. Here, we have defined two $\Theta$-functions:  
\begin{eqnarray}
 \label{30}
&& \Theta_{\vec k, \vec k+\vec p; \vec k', \vec k' -\vec p, \lambda } = \\
 &&   \quad  =  \Theta\big( \lambda -|\varepsilon_{\vec k, \lambda} -  \varepsilon_{\vec k + \vec p, \lambda}
  +  \varepsilon_{\vec k', \lambda} -  \varepsilon_{\vec k' -\vec p, \lambda} |
  \big)\,,  \nonumber \\
&& \Theta_{\vec k, \vec k + \vec p,\lambda} = 
  \Theta\big( \lambda -|\varepsilon_{\vec k, \lambda} -  \varepsilon_{\vec k + \vec p, \lambda}|\big) \, .
  \end{eqnarray}
  They guarantee that only transitions with excitation energies 
smaller than $\lambda$ are kept in $\mathcal H_{f,\lambda}(t)$. 
Finally we use  relation (\ref{9}) to regroup $\mathcal H_{f,\lambda}(t)$:
\begin{equation}
{\mathcal H}_{f, \lambda}(t) = {\mathcal H}^\alpha_{f, \lambda}(t) + {\mathcal H}^\beta_{f, \lambda}(t)
\end{equation}
 with
\begin{eqnarray}
 \label{31}
  {\mathcal H}^\alpha_{f, \lambda}(t) &=&  \frac{U}{2N} \sum_{\vec k \vec k' \vec p \sigma}  
  \Theta_{\vec k, \vec k+\vec p; \vec k', \vec k' -\vec p, \lambda }  \\ 
&&\times :c^\dag_{\vec k \sigma} \,c^{}_{\vec k + \vec p, \sigma}:  \, :c^\dag_{\vec k', -\sigma} \,
c^{}_{\vec k' - \vec p, -\sigma}: \,, \nonumber\\[0.3cm]
 {\mathcal H}^\beta_{f, \lambda}(t) &=&   - {2U} \sum_{\vec k \sigma}  \Big[
\varphi_{\vec k, \vec k +\vec q, \lambda}
 \frac{\sigma}{2} \, c^\dag_{\vec k \sigma} 
c_{\vec k + \vec q, \sigma}  \\
&&+ \varphi_{\vec k, \vec k -\vec q, \lambda}  \frac{\sigma}{2} \, c^\dag_{\vec k \sigma} 
c_{\vec k - \vec q, \sigma}
\Big] \nonumber \, \nonumber .
 \end{eqnarray} 
$  {\mathcal H}^\alpha_{f, \lambda}(t)$ only depends on fluctuation operators, 
whereas $ {\mathcal H}^\beta_{f, \lambda}(t)$ is linear to  the external field via
the expectation value $ \langle s^z_{\vec k', -\vec q} \rangle$.
Here we have defined 
\begin{eqnarray}
\label{32} 
\varphi_{\vec k, \vec k -\vec q, \lambda} &=& \frac{1}{N} \sum_{\vec k'}
 \Big(\Theta_{\vec k, \vec k- \vec q, \vec k', \vec k' + \vec q,\lambda} - \Theta_{\vec k, \vec k-\vec q,\lambda} \Big)
 \langle s^z_{\vec k', -\vec q} \rangle \nonumber  \\
 &=&\varphi_{\vec k, \vec k -\vec q, \lambda}(t) \,.
\end{eqnarray}
Note that also higher-order fluctuation contributions both to $\mathcal H_{f,\lambda}(t)$ and $X_{\lambda, \Delta \lambda}(t)$ could be considered. 
 Their inclusion would extend the range of validity of the present approach, which is restricted to the range from small to intermediate coupling $U/W \lesssim 1$, to larger values.   
However, they would further complicate the evaluation of transformation (\ref{24}) and will be neglected.

Similarly,  from Eq.~(\ref{23})
one finds the following expression for the generator $X_{\lambda, \Delta \lambda}$:
 \begin{equation}
 \label{33}
 X_{\lambda, \Delta \lambda}(t) = X^\alpha_{\lambda, \Delta \lambda}(t)
 + X^\beta_{\lambda, \Delta \lambda}(t) 
 \end{equation}
 with
   \begin{eqnarray}
 \label{34}  
 X^\alpha_{\lambda, \Delta \lambda}(t) &=&   \frac{1}{2N} \sum_{\vec k \vec k' \vec p \sigma}
A_{\vec k, \vec k + \vec p; \vec k', \vec k' -\vec p}(\lambda, \Delta \lambda) \\
&&\times 
: c^\dag_{\vec k \sigma} \,c^{}_{\vec k + \vec p, \sigma} \, c^\dag_{\vec k', -\sigma} \,c^{}_{\vec k' - \vec p, -\sigma}:  \,,
\nonumber \\
X^\beta_{\lambda, \Delta \lambda}(t)
&=& 
- 2\sum_{\vec k \sigma} \Big( 
{B}_{\vec k, \vec k +\vec q}(\lambda, \Delta \lambda) \frac{ \sigma}{2} c^\dag_{\vec k \sigma} 
c_{\vec k + \vec q, \sigma}  
\label{34a}
 \\
&& +
{B}_{\vec k, \vec k -\vec q}(\lambda, \Delta \lambda) \frac{ \sigma}{2} c^\dag_{\vec k \sigma} 
c_{\vec k - \vec q, \sigma}
\Big)  \nonumber \, .
 \end{eqnarray}
The coefficients in Eqs.~(\ref{34}),~(\ref{34a})  are defined by   
     \begin{eqnarray}
 \label{35}
&&A_{\vec k, \vec k + \vec p; \vec k', \vec k' -\vec p}(\lambda, \Delta \lambda) =\\
&&\qquad\qquad   =\frac{\Theta_{\vec k, \vec k+\vec p; \vec k', \vec k' -\vec p} (\lambda, \Delta \lambda) }
{\varepsilon_{\vec k,\lambda}   - \varepsilon_{\vec k + \vec p,\lambda} + \varepsilon_{\vec k',\lambda} 
 - \varepsilon_{\vec k' -\vec p, \lambda} 
 } \, U \nonumber \, . \nonumber
 \end{eqnarray}
 and
 \begin{eqnarray}
 \label{36}
&& {B}_{\vec k, \vec k +\vec q}(\lambda, \Delta \lambda)(t)=\frac{1}{N} \sum_{\vec k'} \Big(A_{\vec k, \vec k + \vec q; \vec k', \vec k' -\vec q}(\lambda, \Delta \lambda)
\nonumber \\
&& \hspace*{3cm}- \hat A_{\vec k, \vec k+ \vec q}(\lambda, \Delta \lambda) \Big)  \langle s^z_{\vec k', \vec q}\rangle \, ,
 \end{eqnarray}
 where
     \begin{eqnarray}
 \label{37}
&&\hat A_{\vec k, \vec k + \vec q}(\lambda, \Delta \lambda) = \frac{\Theta_{\vec k, \vec k+\vec q} (\lambda, \Delta \lambda) }{\varepsilon_{\vec k,\lambda}   - \varepsilon_{\vec k + \vec q,\lambda} } 
  \, U \, . 
 \end{eqnarray}
 In
\begin{eqnarray}
\label{38}
\Theta_{\vec k, \vec k+\vec p; \vec k', \vec k' -\vec p} (\lambda, \Delta \lambda) &=&  
\Theta_{\vec k, \vec k+\vec p; \vec k', \vec k' -\vec p, \lambda}\\
&& \times
\big( 1 - \Theta_{\vec k, \vec k+\vec p; \vec k', \vec k' -\vec p, \lambda - \Delta \lambda} 
 \big) \nonumber 
 \end{eqnarray}
and 
\begin{eqnarray}
 && \Theta_{\vec k, \vec k+\vec q} (\lambda, \Delta \lambda) =  \Theta_{\vec k, \vec k+\vec q, \lambda} 
 \big( 1-  \Theta_{\vec k, \vec k+\vec q, \lambda-  \Delta \lambda}
 \big) 
\end{eqnarray}
the products of $\Theta$-functions assure that only excitations between $\lambda$
and $\lambda- \Delta \lambda$ are eliminated by the unitary transformation (\ref{21}). 

\subsection{Renormalization equations}
 \label{IV.2}
 
Integrating out all transitions induced by $\mathcal H_{f,\lambda}(t)$, the parameters of $\mathcal H_\lambda(t)$ 
will be renormalized.  Only the Coulomb coupling $U$ remains  $\lambda$-independent 
(apart from the $\Theta$-functions in Eq.~(\ref{30})).
The $\lambda$-dependence of the parameters will be derived with  transformation (\ref{24}) 
for an additional step from $\lambda$ to $\lambda - \Delta \lambda$.
The result of  the explicit evaluation has to be compared with the generic expression 
for $ {\mathcal H}_{\lambda - \Delta \lambda}(t)$, which  is obtained by replacing 
$\lambda$ in $\mathcal H_{\lambda}$ [Eqs.~(\ref{26})-(\ref{30})]. In this way one obtains 
the desired renormalization equations, which  
connect the $\lambda$-dependent parameters of $\mathcal H_\lambda$ with those at
cutoff $\lambda - \Delta \lambda$. According to Appendix B we find
 \begin{eqnarray}
 \label{39}
 &&\varepsilon_{\vec k, \lambda -\Delta \lambda} - \varepsilon_{\vec k, \lambda} = 
 \delta\varepsilon^{(1)}_{\vec k, \lambda}  -  \frac{1}{2}  \delta \varepsilon^{(2)}_{\vec k, \lambda} \,,\\
  \label{40}
  &&u_{\vec k, \vec k-\vec q,\lambda -\Delta \lambda}(t) -  
  u_{\vec k, \vec k -\vec q,\lambda}(t)  
  = 
  \sum_{n=1}^4\delta u^{(n)}_{\vec k, \vec k- \vec q, \lambda}(t)   \,,  \\ 
  \label{41}
  &&  v_{\vec k, \vec k +\vec p -\vec q; \vec k', \vec k' -\vec p,  \lambda -\Delta \lambda}(t) -  
  v_{\vec k, \vec k +\vec p -\vec q; \vec k', \vec k' -\vec p,  \lambda}(t)   \nonumber \\
  && \hspace*{2.5cm}  =  
  \sum_{n=1}^3
  \delta v^{(n)}_{\vec k, \vec k +\vec p -\vec q; \vec k', \vec k' -\vec p,  \lambda}(t) 
  \, . 
  \end{eqnarray}
 The renormalization contributions on the right-hand sides of these equations are of order $U$ and $U^2$.
 They  are given 
 in Appendix B: 
  $ \delta\varepsilon^{(1)}_{\vec k, \lambda}$, 
  $\delta  \varepsilon^{(2)}_{\vec k, \lambda}$  by Eqs.~(\ref{B10}) and (\ref{B24});
  $\delta u^{(n)}_{\vec k, \vec q, \lambda}(t)$, $(n=1 \ldots 4)$ by Eqs.~(\ref{B2}), (\ref{B12}), (\ref{B17}), and (\ref{B20});
   and finally 
   $ \delta v^{(n)}_{\vec k, \vec k +\vec p -\vec q; \vec k', \vec k' -\vec p,  \lambda}(t)$, $(n=1,2,3)$,  by 
  Eqs.~(\ref{B3}), (\ref{xyz1}), and (\ref{xyz2}).   In order to reduce the operator structure of 
 $\mathcal H_{\lambda- \Delta \lambda}(t)$ to operators which 
 appear in $\mathcal H_\lambda(t)$ an additional factorization of higher operator 
terms has been performed. Therefore, the expectation values 
  $\langle c_{\vec k \sigma}^\dag  c_{\vec k \sigma }\rangle$
 and $\langle c_{\vec k \sigma}^\dag  c_{\vec k \pm \vec q, \sigma }\rangle$ 
 enter the renormalization contributions.

 The renormalization equations 
 have to be solved numerically, starting from the initial parameters 
 of the original model ${\mathcal H}(t)$, i.e.,
    \begin{eqnarray}
 \label{42}
&& {\varepsilon}_{\vec k,\Lambda} = {\varepsilon}_{\vec k} + U \langle n \rangle/2 \, ,\\
&&u_{\vec k,\vec k-\vec q, \Lambda}(t) = 0\,,\\
 && 
  v_{\vec k, \vec k+\vec p -\vec q; \vec k', \vec k'- \vec p , \Lambda}(t)=0  \, .\label{42a}
 \end{eqnarray}
Suppose, the expectation values on the right-hand side of Eqs.~(\ref{39})-(\ref{41}) are known, the  renormalization procedure from $\Lambda$ to $\lambda=0$ 
  leads to the fully renormalized Hamiltonian 
  $$\tilde{\mathcal H}(t)= 
 \mathcal H_{0,\lambda =0} + \hat{\mathcal H}_{h,\lambda=0}(t) $$ 
 with 
  \begin{eqnarray}
  \label{43}
&&{\mathcal H}_{0,\lambda=0} = \sum_{\vec k \sigma} \tilde{\varepsilon}_{\vec k} \, 
  c^\dag_{\vec k \sigma}  c^{}_{\vec k \sigma}   \,, \\
 \label{44}
&& \hat {\mathcal H}_{h,\lambda=0}(t) =  -  
 \sum_{\vec k \sigma}\Big[\Big(\frac{\hat h_{\vec q}(t)}{2} +\tilde u_{\vec k, \vec k- \vec q}(t)\Big) \, 
 \frac{\sigma}{2} c^\dag_{\vec k \sigma}  c^{}_{\vec k - \vec q, \sigma}  \nonumber \\
 &&  \phantom{ \hat {\mathcal H}_{h,\lambda=0}(t)} \qquad \qquad \qquad+ \textrm{H.c.}\Big] 
    \\
 &&  \phantom{ \hat {\mathcal H}_{h,\lambda=0}(t)}- \frac{1}{N} \sum_{\vec k \vec k' \vec p \sigma} \big[
 \tilde v_{\vec k, \vec k + \vec p -\vec q; \vec k', \vec k' -\vec p}(t) \nonumber \\
 && \phantom{ \hat {\mathcal H}_{h,\lambda=0}(t)}\times  \frac{\sigma}{2}
: c^\dag_{\vec k\sigma} c_{\vec k +\vec p -\vec q, \sigma} c^\dag_{\vec k' , -\sigma}
c_{\vec k' -\vec p, -\sigma}:  + \textrm{H.c.} \big] 
\, . \nonumber
 \end{eqnarray}
 The  tilde symbols denote the fully renormalized quantities at 
 $\lambda=0$. All excitations from $\mathcal H_{f,\lambda}(t)$ have been eliminated, leading to 
 the renormalization of $\mathcal H_{0,\lambda}$ and $\hat{\mathcal H}_{h,\lambda}(t)$. 
 The final Hamiltonian $\tilde{\mathcal H}(t)$ describes a 
 system of free renormalized conduction
 electrons in a renormalized effective field.   Thereby, the  quantities
$\tilde u_{\vec k, \vec k- \vec q}(t)$  and 
$\tilde v_{\vec k, \vec k+\vec p -\vec q; \vec k', \vec k'- \vec p}(t)$ 
depend linearly on  the external field and are time-dependent. This follows from the 
renormalization equations (\ref{39})-(\ref{41}), using the expressions for the renormalization contributions from 
Appendix A, together with the initial conditions (\ref{42}).   Therefore, 
relying on linear response theory with respect to the effective field, any expectation value can be evaluated.

\subsection{Expectation values} 
\label{IV.3}

\subsubsection{Occupation numbers $\langle n_{\vec k\sigma} \rangle$}
\label{IV.3.1}

The yet unknown expectation values on the right-hand side of the renormalization equations
(\ref{39})- (\ref{41}) can  be evaluated self-consistently as follows. Let us first consider 
 the averaged occupation number $\langle n_{\vec k\sigma}\rangle$ 
 for fixed spin $\sigma$. Using Eq.~(\ref{25}), $\langle n_{\vec k\sigma}\rangle$
can be rewritten as
\begin{equation}
\label{45}
\langle n_{\vec k\sigma} \rangle =
\langle c^\dag_{\vec k\sigma}c^{}_{\vec k\sigma}\rangle=\langle \tilde{c}^\dag_{\vec k \sigma}\tilde{c}^{}_{\vec k \sigma}\rangle_{\tilde{\mathcal{H}}}\, .
\end{equation}
In principle, the last expectation value has to be  formed with the time-dependent 
Hamiltonian $\tilde{\mathcal{H}}(t)$. However, restricting ourselves to first order in $h(t)$,  
$\tilde{\mathcal{H}}(t)$ can be replaced by the time-independent Hamiltonian $\tilde{\mathcal H}_{0}$.
$\tilde{c}^{\dag}_{\vec k\sigma}$ is the 
fully renormalized creation operator $\tilde{c}^{\dag}_{\vec k\sigma}=c^{\dag}_{\vec k\sigma}(\lambda\rightarrow 0)$, where $c^{\dag}_{\vec k\sigma}(\lambda)$ is defined by  $c^{\dag}_{\vec k\sigma}(\lambda)
 = e^{X_\lambda} c^\dag_{\vec k \sigma} e^{-X_\lambda}$. 
 
 For $c^{\dag}_{\vec k\sigma}(\lambda)$ an appropriate {\it ansatz} is necessary. We choose 
\begin{eqnarray}
\label{46}
c^\dag_{\vec k \sigma}(\lambda) &=& x^{}_{\mathbf{k},\lambda}c^\dag_{\vec k \sigma}   \\
&&+ \!\frac{1}{2N}\sum_{\vec p\vec k'}
y^{}_{\vec k\vec p\vec k',\lambda}c^\dag_{\mathbf{k-p},\sigma}\!:\!c^\dag_{\vec k',-\sigma}c^{}_{\vec k'-\vec p,-\sigma}\!:
 \, ,  \nonumber
\end{eqnarray}
where the operator structure of (\ref{46}) is again taken over from the lowest 
order expansion in $X_\lambda$ of the unitary transformation. 

In analogy to Eq.~(\ref{21}) renormalization equations for the 
coefficients $x^{}_{\mathbf{k},\lambda}$ and 
$y^{}_{\vec k\vec p\vec k',\lambda}$ can be derived by evaluating  
the renormalization step from $\lambda$ to $\lambda- \Delta \lambda$. One finds
\begin{equation}
\label{47}
y^{}_{\mathbf{kp}\vec k',\lambda-\Delta\lambda}-y^{}_{\mathbf{kp}\vec k',\lambda}=x^{}_{\mathbf{k},\lambda}
A_{\mathbf{k-p},\vec k;\vec k',\vec k'-\vec p}(\lambda,\Delta\lambda) \, . 
\end{equation}
This equation  connects $y^{}_{\mathbf{kp}\vec k',\lambda-\Delta\lambda}$ at cutoff $\lambda -\Delta \lambda$ 
with the coefficients $x^{}_{\mathbf{k},\lambda}$, $y^{}_{\mathbf{kp}\vec k',\lambda}$  
at cutoff $\lambda$. Similarly, also an  
renormalization equation for $x_{\vec k, \lambda}$ can be found. Alternatively one may
start from the anti-commutator relation of $c^\dag_{\vec{k}\sigma}(\lambda)$ and  $c^{}_{\vec{k}\sigma}(\lambda)$.
Taking the expectation value with $\mathcal H_\lambda$, we get
 \begin{equation}
\label{48}
\langle [c^\dag_{\vec{k}\sigma}(\lambda),c^{}_{\vec{k}\sigma}(\lambda)]_+\rangle_{\mathcal{H}_\lambda}=1 \, ,
\end{equation}
or
\begin{eqnarray}
\label{49}
&& |x^{}_{\mathbf{k},\lambda}|^2+\frac{1}{4N^2}\sum_{\mathbf{p}\vec k'}|y^{}_{\vec k\vec p\vec k',\lambda}|^2S^{c}_{\mathbf{k}\mathbf{p}\vec k'}=1 \, .
\end{eqnarray}
Here, a factorization approximation for
\begin{equation}
S^{c}_{\mathbf{kp}\vec k'}=\langle n_{\mathbf{k}-{\vec p}}\rangle(\langle n_{\mathbf{k}'}\rangle-\langle n_{\mathbf{k}'-\vec p}\rangle
+\langle n_{\mathbf{k}'-\vec p}\rangle(2-\langle n_{\mathbf{k}'}\rangle)
\end{equation}
was used. 
Eq.~(\ref{49})  connects the  coefficients $x^{}_{\mathbf{k},\lambda}$ 
and $y^{}_{\mathbf{kp}\vec k',\lambda}$  
for any value of  $\lambda$. 
The equation for $x_{\vec k, \lambda- \Delta \lambda}$ is found
from the sum rule (\ref{49}), when $\lambda$ is replaced by  $\lambda-\Delta\lambda$, i.e.,
\begin{eqnarray}
\label{50}
&& |x^{}_{\mathbf{k},\lambda-\Delta\lambda}|^2=1-\frac{1}{4N^2}\sum_{\mathbf{p}\vec k'}|y^{}_{\mathbf{kp}\vec k',\lambda-\Delta\lambda}|^2S^{c}_{\mathbf{kp}\vec k'}\,.
\end{eqnarray}
Together with Eq.~(\ref{47}) this equation relates $x^{}_{\mathbf{k},\lambda-\Delta\lambda}$ with
 $x^{}_{\mathbf{k},\lambda}$ and $y^{}_{\vec k\vec p\vec k',\lambda}$.  
 Thus Eqs.~(\ref{50}), (\ref{47}) connect the parameter values 
 at $\lambda$ with those at $\lambda -\Delta \lambda$.
Integrating Eqs.~(\ref{47}), (\ref{50}) 
between $\Lambda$ and $\lambda =0$ (thereby using $x^{}_{\mathbf{k},\Lambda}=1$ 
and $y^{}_{\mathbf{kp}\vec k',\Lambda}=0$), we obtain
 \begin{eqnarray}
\label{51}
&& \tilde{c}^\dagger_{\mathbf{k}\sigma}=\tilde{x}^{}_{\mathbf{k}}c^\dagger_{\mathbf{k}\sigma}\\
&& \qquad +\frac{1}{2N}\sum_{\mathbf{p}\vec k'}
\tilde{y}^{}_{\mathbf{kp}\vec k'}c^\dagger_{\mathbf{k-p},\sigma}
:c^\dagger_{\mathbf{k}',-\sigma}c^{}_{\mathbf{k}'-\vec p,-\sigma}:  \, , \nonumber 
\end{eqnarray}
where the tildes again denote the fully renormalized quantities. 
Thus, for $\langle n_{\mathbf k \sigma} \rangle$ the  final result is
\begin{eqnarray}
\label{52}
&& \langle n_{\mathbf{k} \sigma} \rangle=
|\tilde{x}^{}_{\mathbf{k}}|^2f(\tilde{\varepsilon}_{\vec k})  \\ 
&& \quad +\frac{1}{N^2}\sum_{\mathbf{p}\vec k'}\left|
\tilde{y}^{}_{\mathbf{kp}\vec k'}\right|^2
f(\tilde{\varepsilon}_{\vec k-\vec p})f(\tilde{\varepsilon}_{\vec k'})[1-f(\tilde{\varepsilon}_{\vec k'-\vec p})] 
\nonumber  \, ,
\end{eqnarray}
which is independent of $\sigma$ in the paramagnetic state. $f(\tilde{\varepsilon}_{\vec k})$ is the Fermi function.

\subsubsection{Transformation of spin operators}
\label{IV.3.2}

To evaluate the dynamical spin susceptibility we need the transformed
 spin operator, $\tilde s^z_{\vec q} = s^z_{\vec q, \lambda \rightarrow 0}$, where  
$s_{\vec q,\lambda}^z= e^{X_\lambda} s_{\vec q}^z e^{-X_\lambda}$.  For the $\lambda$-dependence
we use the following {\it ansatz}, which corresponds to {\it ansatz} (\ref{28}) for $\hat{\mathcal H}_{h,\lambda}$.  
According to Appendix A we write
  \begin{eqnarray}
 \label{53}
 s^z_{\vec q,\lambda} &=&
\sum_{\vec k \sigma}  \alpha_{\vec k,\vec k -\vec q,\lambda} \, 
\frac{\sigma}{2} c_{\vec k \sigma}^\dag c^{}_{\vec k - \vec q, \sigma}  
 \\
 &&+
   \frac{1}{N} \sum_{\vec k \vec k' \vec p \sigma} 
    \beta_{\vec k,\vec k +\vec p -\vec q, \vec k', \vec k' -\vec p, \lambda}
  \nonumber \\
&&\times   
 \frac{\sigma}{2}
   :c_{\vec k \sigma}^\dag c^{}_{\vec k + \vec p -\vec q, \sigma } 
   c_{\vec k', -\sigma}^\dag c^{}_{\vec k'- \vec p, -\sigma} :  
   \nonumber 
  \end{eqnarray}
with $\lambda$-dependent coefficients $\alpha_{\vec k,\vec k -\vec q,\lambda}$, 
$\beta_{\vec k,\vec k +\vec p -\vec q, \vec k', \vec k' -\vec p, \lambda}$. Their initial values  
 at $\lambda =\Lambda$ are
   \begin{eqnarray}
 \label{54}
 \alpha_{\vec k,\vec k -\vec q,\Lambda} = 1 \, , \quad 
   \beta_{\vec k,\vec k +\vec p -\vec q, \vec k', \vec k' -\vec p, \Lambda} =0 \, .
    \end{eqnarray}
The renormalization equations for the coefficients are found from the transformation step
$s^z_{\vec q, \lambda-\Delta \lambda}= e^{X_{\lambda, \Delta \lambda}}
s^z_{\vec q, \lambda}e^{-X_{\lambda, \Delta \lambda}}$, where in linear response theory 
$X_{\lambda, \Delta \lambda}$ 
can be replaced by its part  $X^{(\alpha)}_{\lambda, \Delta \lambda}$.
A closer inspection shows that the renormalization equations 
can be taken over from  the equations for 
 $u_{\vec k,\vec k-\vec q,\lambda}$ 
 and $ v_{\vec k,\vec k +\vec p -\vec q, \vec k', \vec k' -\vec p, \lambda}$. We get
\begin{eqnarray}
\label{55}
&& \alpha_{\vec k,\vec k -\vec q,\lambda-\Delta \lambda} -\alpha_{\vec k,\vec k -\vec q,\lambda} =
\delta \alpha_{\vec k,\vec k -\vec q,\lambda}^{(1)} \nonumber \\
 &&  \beta_{\vec k,\vec k +\vec p -\vec q, \vec k', \vec k' -\vec p, \lambda- \Delta \lambda} -
  \beta_{\vec k,\vec k +\vec p -\vec q, \vec k', \vec k' -\vec p, \lambda}\ \nonumber \\
  && \qquad \qquad=\delta  \beta_{\vec k,\vec k +\vec p -\vec q, \vec k', \vec k' -\vec p, \lambda}^{(1)}\,,
  \end{eqnarray}
 where
 \begin{eqnarray}
 \label{56}
 && \delta \alpha_{\vec k,\vec k -\vec q, \lambda}^{(1)} = 
 \frac{1}{N} \sum_{\vec k'} \big(\alpha_{\vec k' + \vec q,\vec k', \lambda} 
A_{\vec k, \vec k-\vec q; \vec k', \vec k' +\vec q}(\lambda, \Delta \lambda) \nonumber \\
&& \qquad - \alpha_{\vec k', \vec k' -\vec q, \lambda} 
A_{\vec k, \vec k-\vec q; \vec k' -\vec q, \vec k'}(\lambda, \Delta \lambda)
\big) \langle c_{\vec k',-\sigma}^\dag c_{\vec k', -\sigma}\rangle \nonumber \\
&& \qquad \qquad = -\frac{1}{N} \sum_{\vec k'}  \alpha_{\vec k', \vec k' -\vec q, \lambda} 
A_{\vec k, \vec k-\vec q; \vec k' -\vec q, \vec k'}(\lambda, \Delta \lambda) \nonumber \\
 && \qquad \qquad\times \big(
 \langle c_{\vec k',-\sigma}^\dag c_{\vec k', -\sigma}\rangle
-   \langle c_{\vec k' -\vec q,-\sigma}^\dag c_{\vec k' -\vec q, -\sigma}\rangle
\big)\,,
\end{eqnarray} 
and 
  \begin{eqnarray}
 \label{57}
&&   \delta \beta^{(1)}_{\vec k,\vec k +\vec p -\vec q, \vec k', \vec k' -\vec p, \lambda}= \\
&& \qquad 
\alpha_{\vec k + \vec p,\vec k +\vec p -\vec q,\lambda}
A_{\vec k, \vec k+\vec p; \vec k', \vec k' -\vec p}(\lambda, \Delta \lambda) \nonumber \\
&& \qquad -  \alpha_{\vec k,\vec k -\vec q, \lambda} 
A_{\vec k -\vec q, \vec k+\vec p-\vec q; \vec k', \vec k' -\vec p}(\lambda, \Delta \lambda) \nonumber 
\, .
\end{eqnarray}
Note that in Eq.~(\ref{55}) no 
equivalent to $\delta u^{(2)}_{\vec k, \vec k-\vec q}$ enters, since the latter contribution
was caused by the commutator $[X_{\lambda,\Delta \lambda}, \mathcal H_{f,\lambda}]$ 
(compare Appendix A).    
 Furthermore, renormalizations from the second part in {\it ansatz} (\ref{53}), being proportional to
 $\sim  \beta_{\vec k,\vec k +\vec p -\vec q, \vec k', \vec k' -\vec p, \lambda}$,
 have been neglected. 

Let us finally stress that the renormalization contributions to $\alpha_{\vec k,\vec k-\vec q,\lambda}$ and 
$ \beta_{\vec k,\vec k +\vec p -\vec q, \vec k',\vec k' -\vec p \lambda} $ vanish 
for wave vector $\vec q=0$. This is in accord with the 
rotational invariance of the system in spin space.

\subsection{Dynamical magnetic susceptibility} 
\label{IV.4}

The dynamical $\vec q$- and $\omega$-dependent magnetic susceptibility 
$\chi(\vec q, \omega)$ is defined by the linear response 
 of the averaged spin $\langle s^z_{-q} \rangle(t)$ to the small external field 
 $h(t)$. Since $\hat{\mathcal H}_{h,\lambda=0}$ itself is proportional to $h(t)$: 
  \begin{equation}
 \label{58}
 \langle s^z_{-q} \rangle (t) 
= - i \int_0^\infty dt' 
 \langle [\tilde s_{-\vec q}^z(t') , \hat{ \mathcal H}_{h,\lambda=0}(t- t')] \rangle_{\tilde{\mathcal H}_0} \,.
 \end{equation}
Here we have again used  the unitary invariance (\ref{25}) of operator expressions under a trace.
The expectation value in Eq.~(\ref{58}) is formed with the  renormalized one-particle Hamiltonian
$\tilde{\mathcal H}_0 = \mathcal H_{0,\lambda =0} =  \sum_{\vec k \sigma} 
\tilde{\varepsilon}_{\vec k} c_{\vec k \sigma}^\dag c_{\vec k \sigma}$. Both
$\tilde s_{-\vec q}^z$ and $\hat{ \mathcal H}_{h,\lambda=0}(t)$ are the fully renormalized quantities. They are given by
 \begin{eqnarray}
 \label{59}
 \tilde s^z_{\vec q} &=&   \nonumber 
\sum_{\vec k \sigma}  \tilde \alpha_{\vec k,\vec k -\vec q} \, 
\frac{\sigma}{2} c_{\vec k \sigma}^\dag c^{}_{\vec k - \vec q, \sigma}  
 \\
 &&+
   \frac{1}{N} \sum_{\vec k \vec k' \vec p \sigma} 
    \tilde \beta_{\vec k,\vec k +\vec p -\vec q, \vec k', \vec k' -\vec p}
  \nonumber \\
&&\times    
 \frac{\sigma}{2}
   :c_{\vec k \sigma}^\dag c^{}_{\vec k + \vec p -\vec q, \sigma } 
   c_{\vec k', -\sigma}^\dag c^{}_{\vec k'- \vec p, -\sigma} :  \, ,
   \end{eqnarray}
 where  $ \tilde \alpha_{\vec k,\vec k -\vec q}$ and 
 $ \tilde \beta_{\vec k,\vec k +\vec p -\vec q, \vec k', \vec k' -\vec p}$ are independent of $h(t)$, and
\begin{eqnarray}
   \label{60}
      \hat {\mathcal H}_{h,\lambda=0}(t) &=& -
   \sum_{\vec k \sigma} \Big(\frac{\hat h_{\vec q}(t)}{2}+ \tilde u_{\vec k,\vec k -\vec q}(t)\Big) \, 
\frac{\sigma}{2} c_{\vec k \sigma}^\dag c^{}_{\vec k - \vec q, \sigma}  
 \nonumber\\
 &&-
   \frac{1}{N} \sum_{\vec k \vec k' \vec p \sigma}  
   \tilde v_{\vec k, \vec k+\vec p -\vec q; \vec k', \vec k'- \vec p}(t) 
\nonumber  \\
&&\times   
 \frac{\sigma}{2}
   :c_{\vec k \sigma}^\dag c^{}_{\vec k + \vec p -\vec q, \sigma } 
   c_{\vec k', -\sigma}^\dag c^{}_{\vec k'- \vec p, -\sigma} : 
   \nonumber    \\ 
   &&+ \textrm{H.c.} \,  . \
 \end{eqnarray} 
Thus, Eq.~(\ref{58}) reduces to
\begin{eqnarray}
   \label{61}
&&  \langle s^z_{-q} \rangle (t) =    \sum_{\vec k } 
 \tilde{\alpha}^*_{\vec k, \vec k -\vec q} \, \chi_{\vec k}^0(\vec q, \omega) \,
\Big( \frac{\hat h_{\vec q}(t)}{2}+  \tilde u_{\vec k, \vec k- \vec q}(t) \Big)
 \nonumber \\
 && \qquad\qquad+  \frac{1}{ N^2} \sum_{\vec k \vec k' \vec p } 
  \tilde \beta_{\vec k,\vec k +\vec p -\vec q, \vec k', \vec k' -\vec p}^* \, 
\chi^0_{\vec k, \vec k', \vec p}(\vec q, \omega)  \nonumber \\
&& \hspace*{3cm} \times   \tilde v_{\vec k, \vec k+\vec p -\vec q; \vec k', \vec k'- \vec p}(t) \, ,
\end{eqnarray} 
where we have introduced $\vec k$-resolved susceptibilities
     \begin{eqnarray}
  \label{62}
&& {\chi}^0_{\vec k}(\vec q,\omega) = \nonumber\\
 &&   \frac{i}{2}  \int_0^\infty 
 \langle [  (c_{\vec k\sigma}^\dag c^{}_{\vec k -\vec q,\sigma})^\dag (t'), c_{\vec k\sigma}^\dag c^{}_{\vec k -\vec q,\sigma}  ] 
 \rangle_{\tilde{\mathcal H}_0} 
 \,  e^{i ( \omega + i\eta)t'} \, dt' \nonumber \\
&& \nonumber \\
&&  =  \frac{1}{2}\frac{f(\tilde{\varepsilon}_{\vec k -\vec q}) -f(\tilde{\varepsilon}_{\vec k})}{(\tilde{\varepsilon}_{\vec k} - \tilde{\varepsilon}_{\vec k- \vec q}) -( \omega + i\eta)}  \, ,
\end{eqnarray}
and 
      \begin{eqnarray}
  \label{63}
&& {\chi}^0_{\vec k,\vec k', \vec p}(\vec q,\omega) = \nonumber\\
  && \quad   \frac{i}{2}  \int_0^\infty \langle [
  \big( :c_{\vec k \sigma}^\dag c^{}_{\vec k + \vec p -\vec q, \sigma } 
   c_{\vec k', -\sigma}^\dag c^{}_{\vec k'- \vec p, -\sigma} :)^\dag (t')  \nonumber \\
   &&  \qquad
    :c_{\vec k \sigma}^\dag c^{}_{\vec k + \vec p -\vec q, \sigma } 
   c_{\vec k', -\sigma}^\dag c^{}_{\vec k'- \vec p, -\sigma} :
   ] \rangle_{\tilde{\mathcal H}_0} \, 
   e^{i( \omega + i\eta)t'}    dt' \nonumber \\
   && \nonumber \\
 && \quad =\frac{1}{2}\frac{N_{\vec k \vec k' \vec p}}{{\omega}_{\vec k \vec k' \vec p} -(\omega + i\eta)} 
 \end{eqnarray} 
 with 
  \begin{eqnarray}
  \omega_{\vec k \vec k' \vec p} &=&
  \tilde{\varepsilon}_{\vec k} - \tilde{\varepsilon}_{\vec k +\vec p -\vec q} + \tilde{\varepsilon}_{\vec k'} 
  - \tilde{\varepsilon}_{\vec k' -\vec p} \,,  \\
  N_{\vec k \vec k' \vec p} &=&
f(\tilde{\varepsilon}_{\vec k'- \vec p}) \big[ 1- f(\tilde{\varepsilon}_{\vec k'})\big]
 \big[ f(\tilde{\varepsilon}_{\vec k+ \vec p -\vec q}) - f(\tilde{\varepsilon}_{\vec k}) \big] \nonumber
 \\
  &\!\!\!\!\!\!\!\!+&\!\!\!\!\!\!\!\!f(\tilde{\varepsilon}_{\vec k})\big[1- f(\tilde{\varepsilon}_{\vec k +\vec p -\vec q})\big]
  \big[ f(\tilde{\varepsilon}_{\vec k'- \vec p}) - f(\tilde{\varepsilon}_{\vec k'}) \big] 
  \,  .
   \end{eqnarray}
Note that in both susceptibilities (\ref{62}) and (\ref{63}) renormalized energies 
$\tilde\varepsilon_{\vec k}$ enter and not the unrenormalized energies $\varepsilon_{\vec k}$ as in Eq.~(\ref{19}). 

According to Sect.~\ref{IV.2} and Appendix B the quantities
$\tilde u_{\vec k, \vec k- \vec q}(t)$  and 
$\tilde v_{\vec k, \vec k+\vec p -\vec q; \vec k', \vec k'- \vec p}(t)$ 
depend linearly on the external field $h(t)$ via the $\vec k$-resolved spin operator expectation values 
$\langle s^z_{\vec k, -\vec q}\rangle(t)$. In order to simplify the further calculation 
we shall trace back these quantities to the expectation value $\langle s^z_{-\vec q}\rangle(t)$ of the full spin operator. 
This is done by assuming that the coefficients $ B^{(n)}_{\vec p \bar{\vec p};  \vec k \vec q}(\lambda,\Delta \lambda)$, $(n=2,3,4)$  
and  $D^{(n)}_{\bar{\vec p}; \vec k \vec k' \vec p \vec q}(\lambda, \Delta \lambda)$, $(n=2,3)$ 
defined in Appendix B are almost independent of the wave vector $\bar {\vec p}$. For example, we use for the 
renormalization contribution $ \delta  u_{\vec k, \vec k -\vec q,\lambda}^{(2)}(t)$ [Eq.~(\ref{B12})]:
    \begin{eqnarray}
  \label{64}
&& \delta  u_{\vec k, \vec k -\vec q,\lambda}^{(2)}(t)  \approx 
 - \frac{U}{2N^2} \sum_{\vec p, \bar{\vec p}} 
 B_{\vec p \bar{\vec p}; \vec k \vec q}^{(2)}(\lambda, \Delta \lambda) \frac{\langle s^z_{-\vec q}\rangle(t)}{N} \;\,
\end{eqnarray}
with $\langle s^z_{-\vec q}\rangle= \sum_{\vec k'} \langle s^z_{\vec k', -\vec q}\rangle $. 
Thus, from $u_{\vec k, \vec k- \vec q, \lambda}(t)$  as well as from
$v_{\vec k, \vec k+\vec p -\vec q; \vec k', \vec k'- \vec p, \lambda}(t)$ 
 a common factor $\langle s^z_{-\vec q}\rangle(t)/N$ can be extracted:
     \begin{eqnarray}
  \label{65}
 &&  u_{\vec k, \vec k- \vec q, \lambda}(t) =   u^0_{\vec k, \vec k- \vec q, \lambda} \, 
 \frac{\langle s^z_{-\vec q}\rangle(t)}{N}\,,
  \\
 && v_{\vec k, \vec k+\vec p -\vec q; \vec k', \vec k'- \vec p, \lambda}(t) = 
 v^0_{\vec k, \vec k+\vec p -\vec q; \vec k', \vec k'- \vec p, \lambda} 
 \frac{\langle s^z_{-\vec q}\rangle(t)}{N} \nonumber\, ,\\
  \end{eqnarray}
where we introduced time-independent quantities $ u^0_{\vec k, \vec k- \vec q, \lambda}$ and 
$ v^0_{\vec k, \vec k+\vec p -\vec q; \vec k', \vec k'- \vec p, \lambda}$. 
They obey the following renormalization equations: 
 \begin{eqnarray}
 \label{66}
  && u^0_{\vec k, \vec k-\vec q,\lambda -\Delta \lambda} -  
  u^0_{\vec k, \vec k -\vec q,\lambda} = \delta u^{0(1)}_{\vec k, \vec k- \vec q, \lambda}
+ \sum_{n=2}^4 \delta u^{0(n)}_{\vec k, \vec k- \vec q, \lambda} \,,\nonumber\\ \\ 
   \label{67}
  &&  v^0_{\vec k, \vec k +\vec p -\vec q; \vec k', \vec k' -\vec p,  \lambda -\Delta \lambda} -  
  v^0_{\vec k, \vec k +\vec p -\vec q; \vec k', \vec k' -\vec p,  \lambda}  = \nonumber \\
  && \quad\delta v^{0(1)}_{\vec k, \vec k +\vec p -\vec q; \vec k', \vec k' -\vec p,  \lambda} 
    + \sum_{n=2}^3
  \delta v^{0(n)}_{\vec k, \vec k +\vec p -\vec q; \vec k', \vec k' -\vec p,  \lambda} 
   \, ,
  \end{eqnarray}
  with the time-independent renormalization contributions given in  Appendix C. 
Due  to Eq.~(\ref{42}) the initial conditions for $ u^0_{\vec k, \vec k- \vec q, \lambda}$ and 
$ v^0_{\vec k, \vec k+\vec p -\vec q; \vec k', \vec k'- \vec p, \lambda}$ 
at cutoff $\Lambda$ are
 \begin{eqnarray}
 \label{68}
 && u^0_{\vec k, \vec k -\vec q,\Lambda} = v^0_{\vec k, \vec k +\vec p -\vec q; \vec k', \vec k' -\vec p,  \Lambda} =0 \, . 
 \end{eqnarray}

Having Eq.~(\ref{66}), (\ref{67}) we are in a position to rewrite relation (\ref{61}) in a 
time-independent form:  
\begin{eqnarray}
   \label{69}
&&  1  =   \frac{1}{N} \sum_{\vec k } 
 \tilde{\alpha}^*_{\vec k, \vec k -\vec q} \, \chi_{\vec k}^0(\vec q, \omega) \,
 \big( U + \frac{1}{\chi(\vec q, \omega)}+ \tilde u^0_{\vec k, \vec k- \vec q} \big)
 \nonumber \\
 && \qquad\qquad+  \frac{1}{ N^3} \sum_{\vec k \vec k' \vec p } 
  \tilde \beta_{\vec k,\vec k +\vec p -\vec q, \vec k', \vec k' -\vec p}^* \, 
\chi^0_{\vec k, \vec k', \vec p}(\vec q, \omega)  \nonumber \\
&& \hspace*{3cm} \times   \tilde v^0_{\vec k, \vec k+\vec p -\vec q; \vec k', \vec k'- \vec p} \,,
\end{eqnarray}
where, on the right-hand side, we have used the relation
\begin{equation}
\label{70}
\frac{\hat h_{\vec q}(t)}{2} = \Big(U + \frac{1}{\chi(\vec q, \omega)}\Big) \frac{\langle s^z_{-\vec q}\rangle(t)}{N} \, .
\end{equation}
The  coefficients $\tilde u^0_{\vec k, \vec k- \vec q} $ and 
$ \tilde v^0_{\vec k, \vec k+\vec p -\vec q; \vec k', \vec k'- \vec p} $ 
are again fully renormalized. They are  found by solving the 
renormalization equations (\ref{66}), (\ref{67}) due to the initial conditions (\ref{68}). 
Eq.~(\ref{69}) is an implicit equation for the 
dynamical susceptibility $\chi(\vec q, \omega)$ which also enters the  quantities  
$\delta u_{\vec k,\vec k -\vec q, \lambda}^{0(1)}$ and 
$ \delta v^{0(1)}_{\vec k,\vec k +\vec p -\vec q, \vec k', \vec k' -\vec p, \lambda}$ (see Appendix C). 

Solving for $\chi(\vec q, \omega)$ we obtain our final analytical result:
\begin{eqnarray}
   \label{71}
 \chi(\vec q, \omega) &=& \displaystyle    \frac{\displaystyle \frac{1}{N}\sum_{\vec k } 
 \tilde{\alpha}^*_{\vec k, \vec k -\vec q} \, \chi_{\vec k}^0(\vec q, \omega) }{1- \displaystyle \frac{U}{N}\sum_{\vec k } 
 \tilde{\alpha}^*_{\vec k, \vec k -\vec q} \, \chi_{\vec k}^0(\vec q, \omega)
-\Delta(\vec q, \omega)} \, , 
\end{eqnarray} 
with 
\begin{eqnarray}
\label{72}
&&\Delta(\vec q, \omega)=  \frac{1}{N} \sum_{\vec k } 
 \tilde{\alpha}^*_{\vec k, \vec k -\vec q} \, \chi_{\vec k}^0(\vec q, \omega)\tilde u^0_{\vec k, \vec k- \vec q}\\
&&+ \frac{1}{N^3} \sum_{\vec k \vec k' \vec p } 
  \tilde \beta_{\vec k,\vec k +\vec p -\vec q, \vec k', \vec k' -\vec p}^* \, 
\chi^0_{\vec k, \vec k', \vec p}(\vec q, \omega) \tilde v^0_{\vec k, \vec k+\vec p -\vec q; \vec k', \vec k'- \vec p}\nonumber\, .
\end{eqnarray}
The PRM result (\ref{71}) for the dynamical magnetic 
susceptibility of the Hubbard model represents an extension of  the standard RPA expression. 
Besides the new coefficients $\tilde{\alpha}_{\vec k, \vec k -\vec q}$ in the numerator 
and denominator an extra term $\Delta(\vec q, \omega)$ occurs in the denominator, which 
generalizes the overall shape of an RPA expression. Noteworthy the quantities 
$\tilde u^0_{\vec k, \vec k- \vec q}$ and  
$\tilde v^0_{\vec k, \vec k+\vec p -\vec q; \vec k', \vec k'- \vec p}$ in $\Delta(\vec q, \omega)$  
themselves depend  on $\chi(\vec q, \omega)$ (Appendix \ref{appC}). Thus, $\chi(\vec q, \omega)$ 
has to be solved self-consistently from Eqs.~(\ref{71}), (\ref{72}).  
Expression (\ref{71}) reduces to the standard
RPA results when all renormalization effects are disregarded.  
Then, $\tilde {\alpha}_{\vec k,\vec k -\vec q}$ 
and $ \tilde \beta_{\vec k,\vec k +\vec p -\vec q, \vec k', \vec k' -\vec p}$ keep their original 
values $1$ and $0$, whereas $\tilde u^0_{\vec k, \vec k- \vec q} $ and
$ \tilde v^0_{\vec k, \vec k+\vec p -\vec q; \vec k', \vec k'- \vec p}$ 
remain zero and also $\Delta(\vec q, \omega)$ 
vanishes. Then $\chi(\vec q, \omega)$ becomes $\chi_{RPA}(\vec q, \omega)$
given by Eq.~(\ref{20})  
 with $ \chi_0(\vec q, \omega)$ being the unrenormalized dynamical magnetic susceptibility 
 of free electrons [Eq.~(\ref{19})]. \\

In the next section $\chi(\vec q, \omega)$ will be evaluated numerically. Before, 
let us study the  special case $\vec q=0$. Since the total spin $s^z_{\vec q=0}$ commutes with the total Hamiltonian,
i.e.~$[{\mathcal H}, s^z_{\vec q}]=0$,  the dynamical susceptibility vanishes, which can immediately 
be seen from the general expression (\ref{8}) for $\chi(\vec q=0, \omega)$ ($\omega$ finite). The same conclusion can also be drawn from the PRM formalism. Since the total spin $s^z_{\vec q=0}$ commutes with the Hamiltonian it 
also commutes with the generator $X_{\lambda, \Delta \lambda}$. Therefore, the coefficients in representation (\ref{59}) for $\tilde s^z_{\vec q, \lambda}$  will not be renormalized for $\vec q=0$. Thus, we have    
  $\tilde{\alpha}_{\vec k, \vec k}=1$ and $\tilde{\beta}_{\vec k, \vec k+ \vec p, \vec k', \vec k' -\vec p}=0$.
 Similarly, the coefficients $\tilde u_{\vec k, \vec k -\vec q}^{0}$ and 
 $\tilde v^0_{\vec k, \vec k+ \vec p -\vec q; \vec k', \vec k' -\vec p}$ 
 in expression (\ref{60}) for $\hat{\mathcal H}_{h,\lambda =0}(t)$ also vanish for 
 $\vec q=0$. Thus, the quantity $\Delta (\vec q, \omega)$ in the denominator of Eq.~(\ref{71}) vanishes
 and the susceptibility $\chi(\vec q=0, \omega)$ takes the  standard RPA form
 \begin{eqnarray}
\label{73}
&&  \chi(\vec q=0, \omega) =   \displaystyle \frac{\displaystyle \chi^0(0, \omega) }
 {1-U\chi^0(0, \omega)}\, .
\end{eqnarray}
Here,   according to Eq.~(\ref{62}), the susceptibility 
\begin{equation}
\label{73a}
\chi^0(\vec q, \omega)= \frac{1}{N}\sum_{\vec k} \chi_{\vec k}^0(\vec q, \omega)\,,
\end{equation}
 which contains the renormalized energies $\tilde{\varepsilon}_{\vec k}$ 
 and not the unrenormalized energies  $\varepsilon_{\vec k}$.  
 From Eqs.~(\ref{73}) and (\ref{73a}) one immediately concludes that for ${\vec q}=0$ 
 the real and imaginary parts of $ \chi(\vec q, \omega)$ vanish (for any finite
 $\omega$). Compare also Section \ref{V.3} below. On the other hand, when the limit $\omega\to 0$ is taken first, the imaginary part  of $\chi({\vec q}, \omega)$ vanishes 
 for any $\vec q$. This follows from the analytical properties of ${\rm Im} \chi(\vec q, \omega)$  or from Eqs.~(\ref{71})
 and (\ref{73a}). In contrast, the real part  ${\rm Re}\chi(\vec q, \omega=0)$   stays finite and reduces to the static 
 $\vec q$-dependent susceptibility  
 \begin{eqnarray}
 \label{73b}
 \chi(\vec q) = {\rm Re}\chi(\vec q, \omega=0) \, .
 \end{eqnarray}
In the limit $\vec q \rightarrow 0$,  $\chi(\vec q)$ gives the  uniform static susceptibility.
Last but not least, keep in mind that the renormalization equations derived so far, 
exclusively apply to the paramagnetic phase.

\section{Numerical Results and Discussion}
\label{V}

The set of self-consistency Eqs.~(\ref{39})-(\ref{42a}),~(\ref{71}), and (\ref{72})  has to be solved numerically 
 in momentum space. Due to additional internal $\vec k$-sums in the renormalization contributions, 
 we restrict ourselves to a square lattice with $N=24\times 24$ sites using periodic boundary conditions. 
In contrast, for the non-interacting susceptibility $\chi_{\vec k}^0(\vec q, \omega)$ a larger mesh in momentum space of  $2000 \times 2000$ points
is used close to the Fermi surface due to the large variations in this quantity.
Choosing reasonable initial 
values for the various expectation values, the renormalization  starts from the cutoff $\Lambda$ of the original model and 
proceeds in energy steps $\Delta \lambda=0.5 \bar{t}$ until $\lambda =0$ is reached. Then the expectation values are recalculated.
Convergence is assumed to be achieved if all quantities are determined with a relative error less than 10$^{-5}$.  
 We have convinced ourselves that a larger lattice size as well as a smaller value of $\Delta \lambda$ will not modify the  presented results.
 In what follows we measure all energies in units of $\bar{t}$.
   
   \subsection{Band renormalization}
 \label{V.1}
 
 \begin{figure*}[t]
    \begin{center}
     \includegraphics[angle = -0, width = 0.3\textwidth]{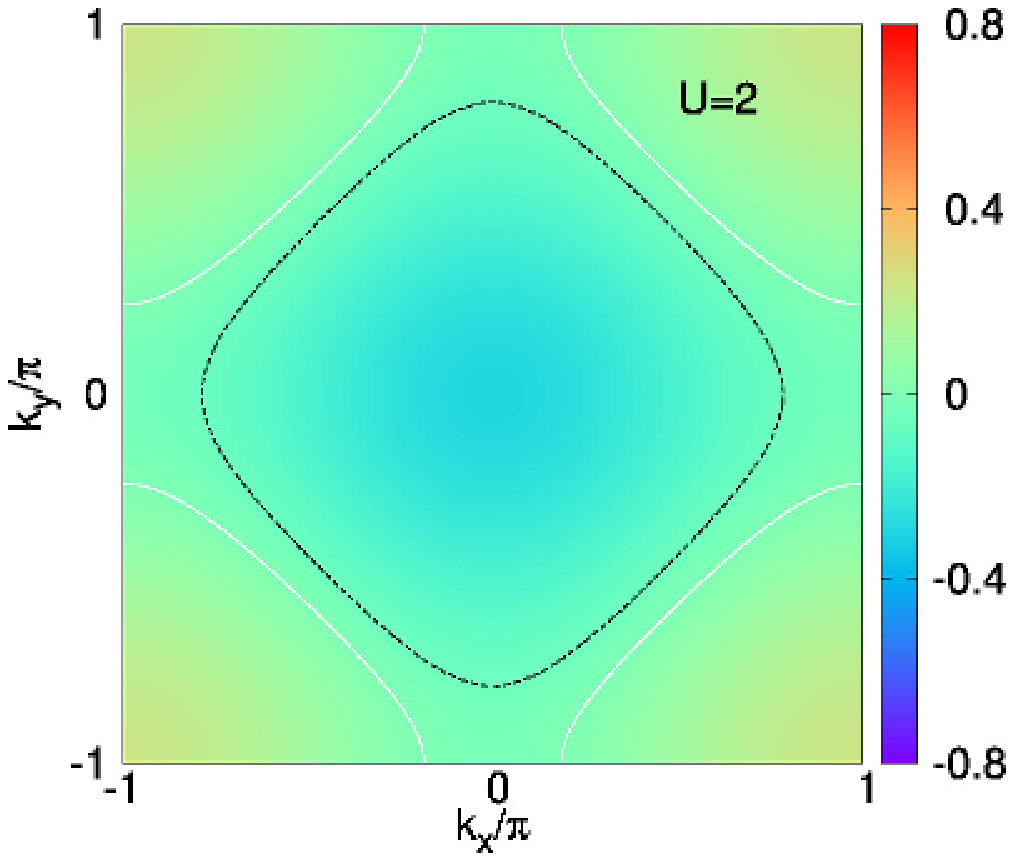}
     \includegraphics[angle = -0, width = 0.3\textwidth]{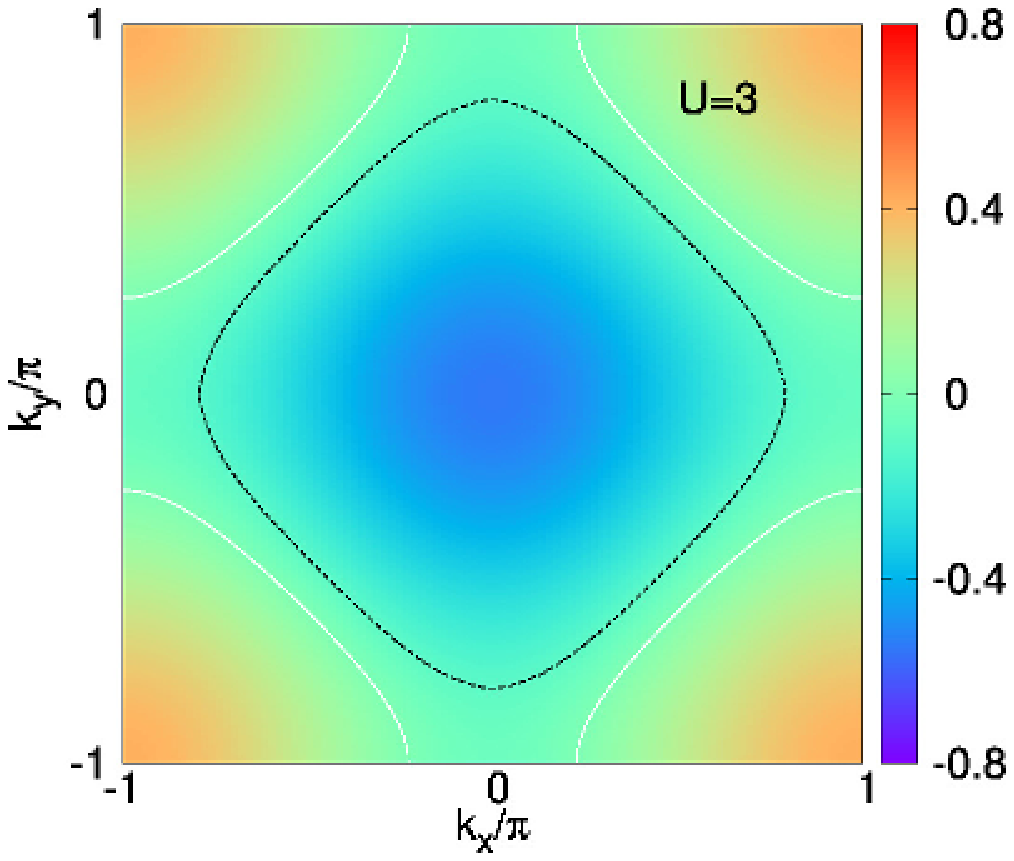}
     \includegraphics[angle = -0, width = 0.3\textwidth]{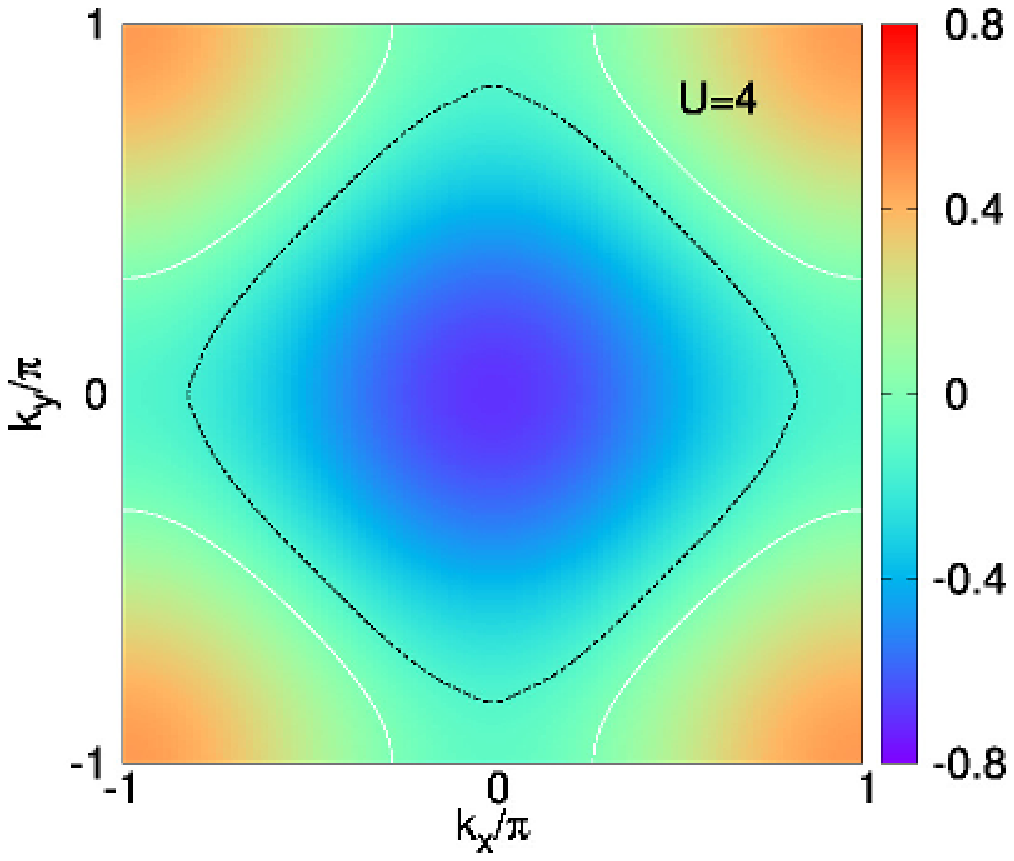}
    \end{center}
\caption{(Color online) Difference between renormalized and unrenormalized one-particle energies in momentum space,
  $\delta\varepsilon_{\vec k}=\varepsilon_{\vec k}-\tilde{\varepsilon}_{\vec k}$, for the 2D Hubbard model with $n=0.9$, see color bar.
  Data obtained by PRM at $T=0$ for  $U=2$ (left panel), $U=3$ (middle panel), and $U=4$ (right panel); 
  the $\delta\varepsilon_{\vec k}=0$-contour is marked in white, the Fermi surface in black.}
\label{fig:9}
\end{figure*}
 
 Correlation effects, which are included in the PRM scheme, lead to a momentum-dependent renormalization $\varepsilon_{\vec k}\to \tilde{\varepsilon}_{\vec k}$ of the band structure
 in the paramagnetic phase. This differs from standard Hartree-Fock~\cite{Pe66,Hi85a}, Gutzwiller~\cite{Gu63,BR70}, or slave-boson treatments~\cite{KR86,DFB92}, where band renormalization either not at all or in terms of a momentum independent band narrowing  takes place. To illustrate the PRM  renormalization of the quasiparticle band, in  
 Fig.~\ref{fig:9} the difference $\delta\varepsilon_{\vec k}=\varepsilon_{\vec k}-\tilde{\varepsilon}_{\vec k}$ is shown for a square lattice Brillouin zone for the particle density $n=0.9$ and three different $U$ values. 
 The overall bandwidth is reduced by a factor of 0.22, 0.34 and 0.45 for  $U=2$, 3 and 4, respectively. 
 Beyond that,  we find that the momentum dependence of $\delta\varepsilon_{\vec k}$ increases with $U$ and is largest near the center ${\vec k}=(0,0)$ and at the corners ${\vec k}=(\pi,\pi)$ of the Brillouin zone. This should have strong impact on 
the uniform and staggered (static) spin susceptibilities which are studied in the next subsection. 

 \subsection{Static spin susceptibility}  
 \label{V.2}

\begin{figure}[t]
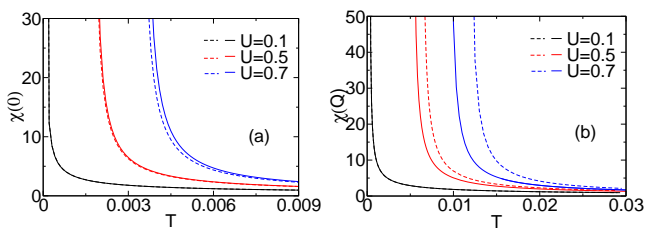

    \begin{center}
     \includegraphics[angle = -0, width = 0.23\textwidth]{Fig2a}
     \includegraphics[angle = -0, width = 0.23\textwidth]{Fig2b}
    \end{center}
 \caption{(Color online) Temperature dependence of the static uniform spin susceptibility $\chi(\vec q=0, T)$ [panel (a)]
and static staggered spin susceptibility $\chi(\vec q= \vec Q,T)$ [panel (b)]  
for the 2D half-filled Hubbard model, evaluated at various 
$U$ by use of the PRM (solid lines) and the RPA (dashed lines). Note the different scales on the $T$ axis.  }
\label{fig:1}       
\end{figure}

\subsubsection{Temperature dependence}
\label{V.2.1}

Let us first consider the half-filled band case $n=1$ and track the behavior of the static susceptibility $\chi(\vec q;T)$ as the temperature $T$ is lowered.  The PRM results for $\chi(\vec q;T)$ are shown in Fig.~\ref{fig:1} for two fixed wave vectors  $\vec{q}=0$ (uniform susceptibility, panel (a)) and  $\vec{q}=\vec{Q}=(\pi,\pi)$ (staggered susceptibility, panel (b)). As one can see, the logarithmic singularity in the density of states at the band center, $\rho(E) \propto \ln (\varepsilon/4\bar{t})$ for $\varepsilon\to 0$, leads to a divergence  of the noninteracting susceptibilities $\chi^0(0;T)\propto -\ln (T/\bar{t})$ and $\chi^0(\vec{Q};T)\propto -[\ln (T/\bar{t})]^2$. This indicates a magnetic instability of the corresponding PRM susceptibilities at some finite $T$ for any $U$. Therefore, since the divergence of $\chi^0(\vec Q; T)$ is stronger than 
that of $\chi^0(0;T)$, the PRM predicts a transition to a magnetic phase with strong antiferromagnetic fluctuations, which  sets in at  a higher temperature.  The larger the $U$-values, the higher are
the transition temperatures.   For very small $U$ the RPA and PRM results are nearly identical.
 As follows from the preceding section, 
 the PRM renormalization of the uniform susceptibility $\chi(0;T)$ is solely caused by the
one-particle energies  $\tilde{\varepsilon}_{\vec k}$ 
(compare Eq.~(\ref{73})), which are barely changed in this limit; cf. Fig.~\ref{fig:9}. On the other hand, the renormalization of $\chi(\vec{Q};T)$ is affected by the coefficients $\tilde{\alpha}_{\bf k, \bf k- {\vec q}} $ as well as 
by the second contribution  $\Delta(\vec{Q})$ in the denominator. This  term shifts the zero of the denominator in Eq.~(\ref{71}),  with the result that 
antiferromagnetic transition temperature is reduced,  as it should be if the correlations/fluctuations of the Hubbard  
system are treated better.  Acceptably  the  antiferromagnetic critical temperature stays larger than the ferromagnetic one.   

 Of course, in 2D the occurrence of a finite transition temperature  
is an artefact of the approximations in the PRM, which is also known from the standard RPA. According to the 
Mermin-Wagner theorem, for    
a 2D model with continuous symmetry long-range order can only occur for $T=0$~\cite{MW66,Hi85a}. 
Indeed, from unbiased numerical approaches~\cite{Hi85a}  it
was shown that  long-range  (antiferromagnetic) order is expected at $T=0$ only for half-filling. Clearly, the PRM  in the present 
version does not overcome this shortcoming. However,  as seen from Fig.~\ref{fig:1}, it gives the  right tendency. One may expect that 
higher order  fluctuation terms, 
not included at present in {\it ansatz} (\ref{27})-(\ref{28}) for $\mathcal H_{\lambda}(t)$ and (\ref{33})-(\ref{34a}) for $X_{\lambda, \Delta \lambda}(t)$ improve the  situation further.

 \subsubsection{$U$-dependence} 
 \label{V.2.2}

\begin{figure}[h]
   \begin{center}
     \includegraphics[angle = -0, width = 0.47\textwidth]{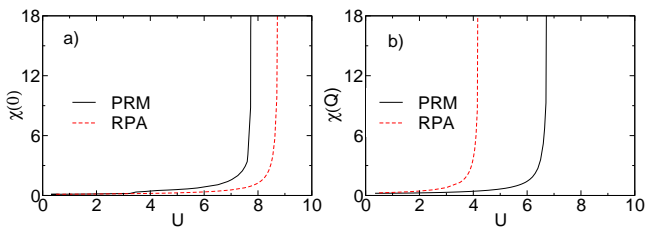}
    \end{center}
\caption{(Color online) $U$-dependence of (a)  the uniform spin susceptibility $\chi(0)$ 
and (b) the static staggered spin susceptibility $\chi(\vec Q)$ at zero temperature  for  the 2D Hubbard model with $n=0.9$, evaluated  by the PRM (solid lines) and the RPA (dashed lines). }
\label{fig:2}
\end{figure}

Figure~\ref{fig:2} shows how (a) the uniform and (b) the staggered spin susceptibilities vary  in the 
paramagnetic phase as the Hubbard interaction $U$ is enhanced at zero temperature. Now we are off half-filling, $n=0.9$.  We again find divergencies in both susceptibilities, signaling a tremendous increase of ferromagnetic and antiferromagnetic correlations.  
If compared to the RPA results (dashed lines), the PRM (solid lines) 
shifts the critical value of $U$ to lower (higher) values in the former (latter) case. 
This is easily understood since the renormalization of $\chi(\vec{q})$ at $\vec{q}=0$ comes (solely) from the PRM 
band narrowing yielding a higher density of states ($\chi^0(0)$) and consequently a lower $U_c$  than within an RPA
treatment. On the other hand, for $\chi(\vec{Q})$ the term $\Delta(\vec{Q})>0$ is more important, 
which enhances $U_c$, i.e.,
suppresses the range where the state with long-range antiferromagnetic correlations pops up. 

\subsubsection{Magnetic phase diagram}
\label{V.2.3}

\begin{figure}
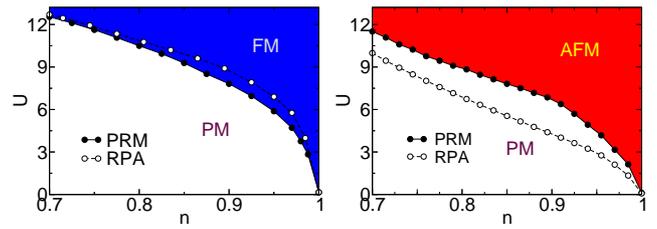

   \begin{center}
     \includegraphics[angle = -0, width = 0.23\textwidth]{Fig4a}
     \includegraphics[angle = -0, width = 0.23\textwidth]{Fig4b}     
    \end{center}
\caption{(Color online)  Zero-temperature phase boundaries between the paramagnetic (PM)
 and ferromagnetic (FM) states (left) respectively
paramagnetic and antiferromagnetic (AFM)  states (right) of the 2D Hubbard model. Filled (open) symbols
mark the transition points in the $n$-$U$ plane obtained from the divergence 
of the corresponding susceptibilities in the PRM (RPA) framework. }
\label{fig:4}
\end{figure}

Tracing the divergencies of the uniform and staggered susceptibilities 
in the $n$-$U$ model-parameter plane at $T=0$, a ground-state phase 
diagram of the 2D Hubbard model can be derived. Thereby,  only paramagnetic states  
with increasingly strong ferromagnetic and antiferromagnetic correlations can be detected. The such kind 
determined phase diagrams agree with those obtained from the corresponding 
order parameter self-consistency equations~\cite{Mah00,DFKTI94,ZIBF11}.    
Of course, the paramagnetic-ferromagnetic and para\-magnetic-antiferromagnetic
phase boundaries have to be determined separately. This has been undertaken 
in Fig.~\ref{fig:4}. Since our PRM treatment of the Hubbard model is a 
weak-to-inter\-mediate 
coupling approach, the calculations were restricted to densities not too far away from
half filling (the instabilities appear at large values of $U$ otherwise).  In the whole
density range studied ($0.7\leq n\leq 1$), the antiferromagnetic instability sets in first,
i.e. an antiferromagnetic state is established before ferromagnetic order can be established.
This corroborates previous Hartree-Fock, RPA, and slave-boson results~\cite{Pe66,KR86,DFB92}.
Quantitative deviations from the RPA phase boundaries exist however (cf., in Fig.~\ref{fig:4}, 
the corresponding transition lines). A tricritical point, where the ferromagnetic and antiferromagnetic
instabilities intersect, is expected to appear at a density slightly smaller than $n=0.7$.
It is not our aim, however, to map out the phase diagram in more detail; simply because 
it is now commonly accepted---owing to numerical studies but not rigorously proven---that long-range ordered phases 
will not be stable in the 2D Hubbard model away from half filling. Thus, as noticed long-time ago,
it appears  that approximative solutions to the simple 2D Hubbard model might do better in 
describing the magnetic features of real quasi-2D materials than the (still not available)
exact solution~\cite{Hi85a}.  In this respect primarily meaningful will be the improved treatment of correlations by the 
PRM in the  paramagnetic phase. 

 \subsection{Dynamic spin susceptibility} 
 \label{V.3}

\begin{figure}
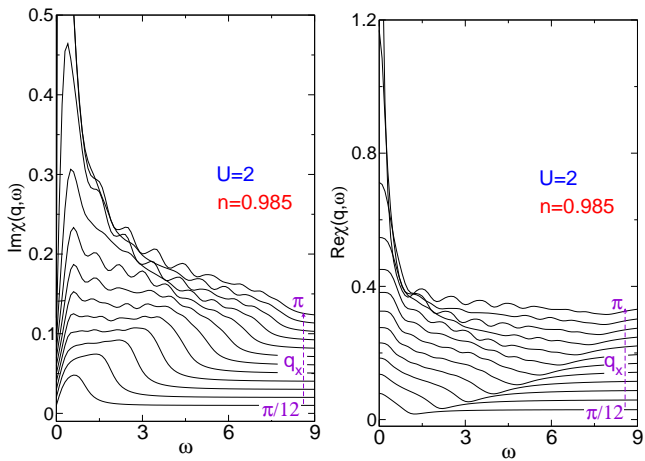

   \begin{center}
     \includegraphics[angle = -0, width = 0.23\textwidth]{Fig5a}
     \includegraphics[angle = -0, width = 0.23\textwidth]{Fig5b}     
    \end{center}
 \caption{PRM results for the imaginary (left panel)  and real (right panel) parts of $\chi({\vec q}, \omega)$ 
 (in arbitrary units) vs frequency at $T=0$ along the 
diagonal  direction  ${\vec q}=(q_x,q_x)$ in the Brillouin zone for  several $q_x$ values 
between $0$ and $\pi$ (see text).  The density is  $n=0.985$, i.e.~close to half-filling, and  $U=2$.  
In  both parts of $\chi({\vec q},\omega)$ 
a strong paramagnon excitation around $\omega=0$ evolves  at the antiferromagnetic wave vector ${\vec Q}=(\pi, \pi)$. Since no enhancement is found in  ${\rm Re}\chi({\vec q},\omega)$ 
 at the ferromagnetic wave vector   
${\vec q}=0$ one concludes that antiferromagnetic are stronger than ferromagnetic  correlations.}
  \label{fig:5}
\end{figure}

The phase diagram Fig.~\ref{fig:4} from Sec.~\ref{V.2} shows 
that transitions from the PM to the AFM 
and from the PM to the FM states approach each other, 
when the density $n$ gets close to half-filling ($n=1$). Thereby the respective critical $U$ values 
go to zero.  In Fig.~\ref{fig:5}  the PRM results for the imaginary and real parts of $\chi({\vec q}, \omega)$ 
are displayed for the density $n=0.985$ very close to 1  and a small coupling $U=2$. Curves are 
shown for  different ${\vec q}$ values along the diagonal direction ${\vec q}= (q_x, q_x)$ in the Brillouin zone. 
The steps between  subsequent $q_x$ curves are chosen as $\pi/12$, where the lowest  $q_x$  
value is $\pi/12$.  Note that  ${\rm Im}\chi({\vec q},\omega)$ for $q_x =0$ (${\vec q}=0$)
vanishes due to rotational symmetry of the total spin density $s^z_{\vec q=0}$.    
A strong 
 paramagnon peak structure is found in ${\rm Im}\chi({\vec q}, \omega)$ at $\omega=0$ around the antiferromagnetic wave vector ${\vec Q}=(\pi, \pi)$. Also in the real part  ${\rm Re}\chi({\vec q}, \omega)$ 
 a  strong peak structure appears for  the same 
 ${\vec q}$ and $\omega$ values. However, no such strong structure is found 
 at small $\vec q$. Since ${\rm Re}\chi({\vec q}, \omega)$ agrees  for $\omega=0$ with the static 
 susceptibility $\chi({\vec q}) = \chi({\vec q}, \omega=0)$  one concludes that antiferromagnetic fluctuations at 
 ${\vec Q}$ dominate ferromagnetic  fluctuations (with ${\vec q} \ll 1)$ 
 already for small deviations from half-filling. \\ 

 %
 
\begin{figure}
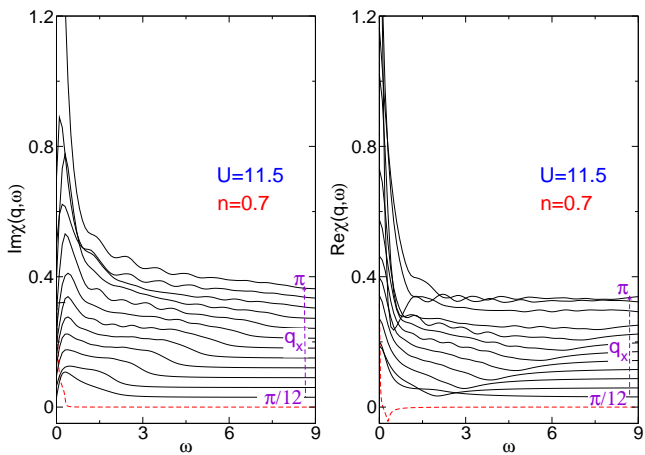

   \begin{center}
     \includegraphics[angle = -0, width = 0.23\textwidth]{Fig6a}
     \includegraphics[angle = -0, width = 0.23\textwidth]{Fig6b}     
    \end{center}
 \caption{PRM result for the same quantities as in Fig.~\ref{fig:5} for the density $n=0.7$ and 
$U=11.5$.   This $U$ is expected to be slightly below the $U$ value at the 
tricritical point where the phase boundaries of the PM-AFM and PM-FM transitions intersect. 
Again, in both parts of $\chi({\vec q},\omega)$ a paramagnon structure is found for the antiferromagnetic 
wave vector ${\vec Q}$ for $\omega \approx 0$. Note that the
 contribution to ${\rm Re}\chi({\vec q},\omega)$ at $\omega=0$ 
 from the smallest $\vec q$  is now enhanced compared to that in
 Fig.~\ref{fig:5}. The evaluation of $\chi(\vec q \rightarrow 0)= {\rm Re}\chi({\vec q \rightarrow 0},\omega=0)$ 
 shows that this quantity rapidly increases 
 by slightly increasing $U$ from 11.5 to the value $U^{PM-FM}_{crit} \simeq 12.5$, where $\chi(q=0)$ would 
 diverge. The red dashed curves result from the standard RPA for an extremely small $\vec q=(0.01, 0.01)$.
 }
  \label{fig:6}
\end{figure}

 Next, let us discuss the circumstances at the density $n=0.7$, which is slighter above the 
 density where the ferromagnetic and antiferromagnetic instabilities are expected to intersect (compare Fig.~\ref{fig:4}). 
 Fig.~\ref{fig:6} shows again the real and imaginary part of $\chi({\vec q}, \omega)$, now
 for $U=11.5$, which is approximately the critical $U$ value for the PM-AFM transition.  
 Indeed, for the antiferromagnetic wave vector
 ${\vec Q}$ there is again a paramagnon structure at $\omega \approx 0$  in $\chi({\vec q},  \omega)$. 
 However, as shown  in the real part of $\chi({\vec q},\omega )$, also 
 ferromagnetic fluctuations (for ${\vec q} \ll 1$ and $\omega=0$) are considerably enhanced 
 compared to the case of Fig.~\ref{fig:5}.   From the $U$ dependence of the uniform static susceptibility  $\chi({\vec q}=0)$ (not shown), 
 one finds that it  tremendously increases 
 for slightly increasing $U$ and diverges at the critical value $U_{crit}^{PM-FM} \approx 12.5$. This divergence would  
 correspond to a transition to a  ferromagnetic phase, if it had not been before  the transition to the antiferromagnetic 
 phase. The red dashed curves in both panels of Fig.~\ref{fig:6} result from a standard RPA calculation for an extremely 
 small $\vec q$ value, $\vec q= (0.01, 0.01)$, which is expected to almost agree  with PRM results.  
 Due to the finite lattice of $24 \times 24$ sites used in the PRM calculation the PRM curves in the figures are restricted to not too
 small $\vec q$ values.

\begin{figure}
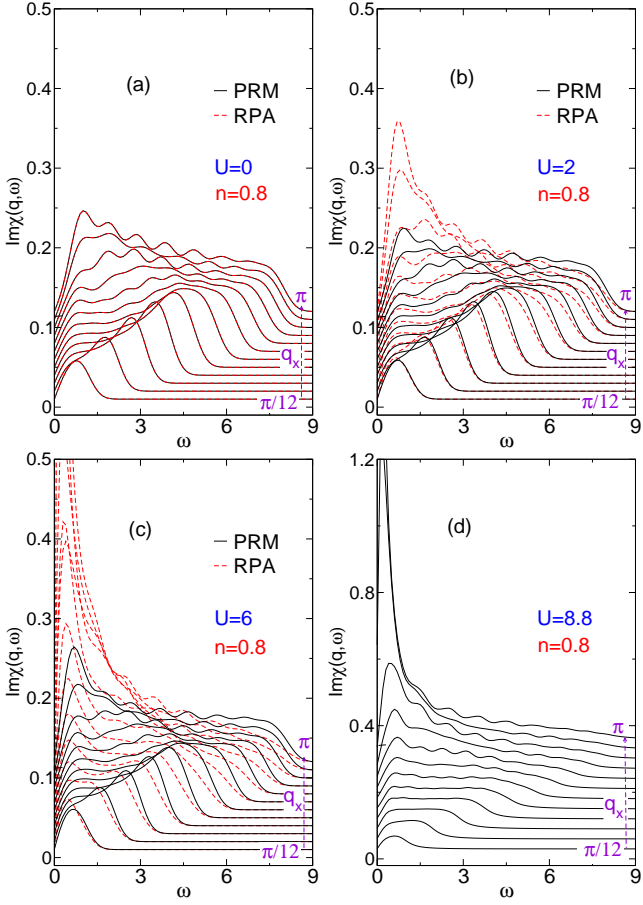

   \begin{center}
     \includegraphics[angle = -0, width = 0.23\textwidth]{Fig7a}
     \includegraphics[angle = -0, width = 0.23\textwidth]{Fig7b}     
     \includegraphics[angle = -0, width = 0.23\textwidth]{Fig7c}
     \includegraphics[angle = -0, width = 0.23\textwidth]{Fig7d}     
    \end{center}
 \caption{Imaginary part of $\chi({\vec q}, \omega)$ (in arbitrary units) vs frequency at $T=0$ 
 for the intermediate density $n=0.8$ along the 
diagonal  direction  ${\vec q}=(q_x,q_x)$ in the Brillouin zone for  $q_x$ values between 
$0$ and $\pi$, evaluated  in the paramagnetic phase. In the first three panels (a)-(c) both the  
PRM results (black solid lines) and the standard RPA results (red dashed lines) are shown, where the 
$U$ values are (a) $U=0$,  (b) $U=2$, and (c) $U=6$.  Note that a paramagnon peak at $\omega=0$ shows up in the RPA result of panel (c) when $q_x$ approaches $\pi$.  
 This finding is consistent with the phase diagram in Fig~\ref{fig:4}, from which the critical $U$ value in the RPA
 for the transition to the antiferromagnetic phase can be extracted as  $U^{RPA}_{crit} \simeq 6.75$.  
 The critical $U$ value in the PRM is higher and amounts to $U^{PRM}_{crit} \simeq 8.8$. This 
 explains that the paramagnon peak is not pronounced in the PRM curves of panels (a)-(c) but in panel (d) 
 for $U=8.8$.
  }
    \label{fig:7}
\end{figure}

Finally, let us compare the PRM with the standard RPA.  
As already discussed in section \ref{II} the RPA arises when all 
renormalization effects are neglected. Panels (a)-(c) of Fig.~\ref{fig:7} show the 
imaginary part ${\rm Im}\chi({\vec q},\omega)$  as a function of $\omega$  for an intermediate 
density $n=0.8$ and three different $U$ values, (a) $U=0$, (b) $U=2$, and (c) $U=6$. 
When ${\vec q}\to {\vec Q}$ the RPA curves (red dashed curves) for $U=6$ exhibit 
a relatively narrow peak at low frequencies, which is again interpreted as paramagnon peak. In contrast, 
the PRM curves at the same $\vec q$ and $U$ are much less pronounced. The different behavior 
is easily understood from the phase diagram in Fig.~\ref{fig:4},
 %
since $U=6$ is much closer to the critical RPA
value $U_{crit}^{RPA}\simeq 6.75$ than to 
the critical value $U^{PRM}_{crit}\simeq 8.8$ from the PRM approach. The corresponding PRM result for 
$U=8.8$ is shown in panel (d) which clearly shows a pronounced paramagnon peak as expected.

In all panels (a) to (d)  the curves  at  ${\vec Q}= (\pi, \pi)$
are more pronounced around $\omega=0$ than those for $\vec q$ values close to  the center of the Brillouin zone. 
From this feature one should not draw the conclusion 
that antiferromagnetic fluctuations are always more important than 
ferromagnetic ones. As was already mentioned, for $\omega$ finite
${\rm Im}\chi({\vec q}, \omega)$ always vanishes  at $\vec q=0$, i.e.~at the ferromagnetic wave vector.
Therefore, as was done in section \ref{IV}, a comparison of antiferromagnetic 
and ferromagnetic fluctuations can only be drawn from the values of the real part $ {\rm Re}\chi({\vec q}, \omega)$ 
of the dynamical susceptibility at $\omega=0$, which is equivalent to the  static $\vec q$ dependent susceptibility  
$\chi(\vec q)= {\rm Re}\chi({\vec q}, \omega=0)$.

\section{Summary}
\label{VI}

Combining linear response theory with the projector-based renormalization method, we presented
a theoretical approach for the evaluations of the susceptibilities that generalizes the standard RPA 
scheme. In this way important  many-body correlations beyond the RPA level were included. To exemplify the advancement the theory was applied to 
the two-dimensional  paradigmatic Hubbard model,  for which an analytical expression for  the 
dynamical spin susceptibility $\chi({\vec q},\omega)$ was derived that improves the RPA result. 
While the uniform spin susceptibility, where ${\vec q}=0$,  still exhibits the standard RPA form, 
renormalized quasiparticle energies enter the band susceptibility contributions $\chi^0({\vec q},\omega)$. At any finite wave-vector, however, the shape of $\chi({\vec q},\omega)$ changes. Besides momentum-dependent prefactors in $\chi^0({\vec q},\omega)$, an extra term occurs 
in the denominator of $\chi({\vec q},\omega)$, which has to be determined self-consistently. This term  particularly changes the pole structure of $\chi({\vec q},\omega)$ in the limit $\omega\to 0$. 
As a result, the magnetic phase diagram, which can be derived from the instabilities of the static spin susceptibilities
at wave-vectors ${\vec q}=0$ or ${\vec q}={\vec Q}$, is modified quantitatively. The same holds for the paramagnon spectrum obtained from the
dynamical response.  Most notably, the better (PRM) treatment of the Coulomb interaction effects severely reduce  
the  exaggeration of the paramagnons in comparison to the RPA, i.e., the  tendency of the Hubbard system  
to develop long-range ferromagnetic or antiferromagnetic order at certain band fillings is weakened.

\section*{Acknowledgements}
The authors would like to thank S. Sykora for valuable discussions.
This research was  funded by Vietnam National Foundation
for Science and Technology Development under Grant No. 103.02-2012.52 and by Deutsche 
Forschungsgemeinschaft through SFB 652, B5.


\begin{appendix}


\section{$\lambda$-dependence of operator quantities}
\label{appA} 

As an example let us justify the {\it ansatz} (\ref{27}) for $\hat {\mathcal H}_{h,\lambda}(t)$. We consider 
the transformation (\ref{21}) for $\hat{\mathcal H}_{h}(t)$ for a small step from 
the original cutoff $\Lambda$ to the somewhat reduced  cutoff $\Lambda - \Delta \lambda$, 
\begin{eqnarray}
\label{A1}
\hat{\mathcal H}_{h,\Lambda - \Delta \lambda}&=& e^{X_{\Lambda, \Delta \lambda}}\, 
 \hat{\mathcal H}_{h}\, e^{-X_{\Lambda, \Delta \lambda}}  \\
&=&   \hat{\mathcal H}_{h} + [X_{\Lambda, \Delta \lambda},   \hat{\mathcal H}_{h}] 
+ \cdots \nonumber
\end{eqnarray}
 Evaluating the  commutator in Eq.~(\ref{A1}) and extracting the one-particle and two-particle 
 contributions one is immediately led to the following structure
\begin{eqnarray}
    \label{A2}
  &&  \hat{\mathcal H}_{h,\lambda}(t) = \\
   && \quad -
   \sum_{\vec k \sigma}  \Big[\Big( \frac{\hat h_{\vec q}(t)}{2}  +
   u_{\vec k,\vec k -\vec q,\lambda}(t)\Big)  \, 
\frac{\sigma}{2} c_{\vec k \sigma}^\dag c^{}_{\vec k - \vec q, \sigma}  + \textrm{H.c.} \Big] \nonumber
 \\
 && \quad -
   \frac{1}{N} \sum_{\vec k \vec k' \vec p \sigma}  \big[
   v_{\vec k, \vec k+\vec p -\vec q; \vec k', \vec k'- \vec p , \lambda}(t) 
  \nonumber \\
&& \quad \times   
 \frac{\sigma}{2}
   :c_{\vec k \sigma}^\dag c^{}_{\vec k + \vec p -\vec q, \sigma } 
   c_{\vec k', -\sigma}^\dag c^{}_{\vec k'- \vec p, -\sigma} : + \textrm{H.c.} \big]
   \nonumber  \, ,
  \end{eqnarray}
  where the prefactors 
  $u_{\vec k, \vec k -\vec q,\lambda}(t)$ and 
   $ v_{\vec k, \vec k+\vec p -\vec q; \vec k', \vec k'- \vec p , \lambda}(t)$ 
   depend on $\lambda$ and also on the external field $h(t)$.

Similarly, for an {\it ansatz} of the transformed spin $s^z_{\vec q, \lambda}$ one starts from 
\begin{eqnarray}
\label{A3}
  s^z_{\vec q,\Lambda- \Delta \lambda}&=& e^{X_{\Lambda, \Delta \lambda}}  s^z_{\vec q,\Lambda}
  e^{- X_{\Lambda, \Delta \lambda}}  \\
  &=&  s^z_{\vec q,\Lambda} + [X_{\Lambda, \Delta \lambda},  s^z_{\vec q,\Lambda}] 
  + \cdots\,, \nonumber
  \end{eqnarray}
 where $s^z_{\vec q,\Lambda}= s^z_{\vec q}$.  
 In lowest order perturbation theory one obtains
  \begin{eqnarray}
  \label{A4}
  s^z_{\vec q,\Lambda- \Delta \lambda} &=&
   \sum_{\vec k \sigma} \frac{\sigma}{2} \, 
 c^\dag_{\vec k \sigma} \, c^{}_{\vec k -\vec q, \sigma}     \\
 && +  
    \frac{1}{2N} \sum_{\vec k \vec k' \vec p }
A^{}_{\vec k, \vec k +\vec p; \vec k', \vec k' -\vec p} (\Lambda, \Delta \lambda)
   \nonumber \\
  && \times    \sum_{\sigma} \frac{\sigma}{2} \Big(
 \, c^\dag_{\vec k, \sigma}\, c^{}_{\vec k + \vec p -\vec q, \sigma} \, 
  :c^\dag_{\vec k',- \sigma}\, c^{}_{\vec k' -\vec p,-\sigma}:   \nonumber \\
    && - 
  c^\dag_{\vec k+ \vec q, \sigma}\, c^{}_{\vec k +\vec p,\sigma} \, 
   :c^\dag_{\vec k', -\sigma}\, c^{}_{\vec k' - \vec p, -\sigma}:   \nonumber \\
&&  +  
   :c^\dag_{\vec k, -\sigma}\, c^{}_{\vec k +\vec p, -\sigma}: \, 
  c^\dag_{\vec k', \sigma}\, c^{}_{\vec k' -\vec p -\vec q,\sigma} \nonumber \\
    && -  
     :c^\dag_{\vec k, -\sigma}\, c^{}_{\vec k + \vec p,-\sigma}: \, 
   c^\dag_{\vec k' +\vec p, \sigma}\, c^{}_{\vec k' - \vec p, \sigma} 
  \Big)   \nonumber 
\end{eqnarray}
which fulfills $s_{\vec q,\lambda}^z= (s_{-\vec q, \lambda}^z)^\dag$. Thus, one arrives at 
   \begin{eqnarray}
 \label{A5}
 s^{z}_{\vec q,\lambda} &=&
   \sum_{\vec k \sigma} \alpha_{\vec k,\vec k -\vec q, \lambda}  \, \frac{\sigma}{2} 
 c^\dag_{\vec k \sigma} \, c^{}_{\vec k -\vec q, \sigma}    \\
 &&+
\frac{1}{N} \sum_{\vec k \sigma} \beta_{\vec k,\vec k +\vec p -\vec q, \vec k',\vec k' -\vec p \lambda} 
 \nonumber  \\  
  &&\times 
  \frac{\sigma}{2} 
  :c^\dag_{\vec k, \sigma}\, c^{}_{\vec k +\vec p -\vec q, \sigma} \, 
  c^\dag_{\vec k', -\sigma}\, c^{}_{\vec k' -\vec p,-\sigma} :\,, \nonumber 
   \end{eqnarray}
where again the one-particle and two-particle contributions were extracted.  
This result agrees with Eq.~(\ref{53}). 

\section{Evaluation of commutators}
\label{appB}

In this appendix we evaluate the commutators from transformation (\ref{24}), which are responsible for 
the renormalization of $\mathcal H_\lambda(t)$.

\subsection{Commutator $ [X_{\lambda, \Delta \lambda}(t), {\mathcal H}_{h, \lambda}(t) ]$}
\label{appB.1)}

Since $\mathcal H_{h,\lambda}(t)$ is linear in the external field, the generator  $X_{\lambda, \Delta \lambda}(t)$ 
can be limited to the part  $X^\alpha_{\lambda, \Delta \lambda}(t)$ in Eq.~(\ref{33}).  
For the part of $\hat{\mathcal H}_{h, \lambda}(t)$ proportional to $u$ , we find 
   \begin{eqnarray}
 \label{B1}
 && [X^\alpha_{\lambda, \Delta \lambda}(t),  \hat{\mathcal H}_{h, \lambda}(t)]\big|_u   = \\
 && \quad - \sum_{\vec k\sigma} \Big\{
\Big(\frac{\hat h_{\vec q}(t)}{2}+u_{\vec k,\vec k-\vec q, \lambda}\Big) \nonumber \\
&& \quad \times
[X^\alpha_{\lambda, \Delta \lambda}(t), \frac{\sigma}{2} c^\dag_{\vec k \sigma} c^{}_{\vec k -\vec q,\sigma} ] + \textrm{H.c.} \Big\}
\nonumber \\
&& = -\sum_{\vec k \sigma} \Big\{
\Big(\frac{\hat h_{\vec q}(t)}{2}+ \delta u_{\vec k, \vec k -\vec q, \lambda}^{(1)}\Big) \frac{\sigma}{2}
c_{\vec k \sigma}^\dag c^{}_{\vec k -\vec q, \sigma} + \textrm{H.c.} \Big\}
\nonumber \\
&& \quad -   \frac{1}{N} \sum_{\vec k \vec k' \vec p \sigma} \Big\{
   \delta v^{(1)}_{\vec k,\vec k +\vec p -\vec q, \vec k', \vec k' -\vec p, \lambda}
  \nonumber \\
&& \quad \times   
 \frac{\sigma}{2}
   :c_{\vec k \sigma}^\dag c^{}_{\vec k + \vec p -\vec q, \sigma } 
   c_{\vec k', -\sigma}^\dag c^{}_{\vec k'- \vec p, -\sigma} :  + \textrm{H.c.} \Big\}
   \nonumber \, . \\
   && \nonumber
\end{eqnarray} 
Here the factors $\delta u_{\vec k, \vec k -\vec q, \lambda}^{(1)}$, and 
$\delta v^{(1)}_{\vec k,\vec k +\vec p -\vec q, \vec k', \vec k' -\vec p, \lambda}$ 
are given by
 \begin{eqnarray}
 \label{B2}
&&  \delta u_{\vec k,\vec k -\vec q, \lambda}^{(1)} =  \\
&& - \frac{1}{N} \sum_{\vec k'} \Big(\frac{\hat h_{\vec q}(t)}{2}+ u_{\vec k', \vec k' -\vec q, \lambda} \Big)
A_{\vec k, \vec k-\vec q; \vec k' -\vec q, \vec k'}(\lambda, \Delta \lambda) \nonumber \\
 && \times \big(
 \langle c_{\vec k',-\sigma}^\dag c^{}_{\vec k', -\sigma}\rangle
-   \langle c_{\vec k' -\vec q,-\sigma}^\dag c^{}_{\vec k' -\vec q, -\sigma}\rangle
\big) \nonumber
\end{eqnarray} 
and
  \begin{eqnarray}
 \label{B3}
&&   \delta v^{(1)}_{\vec k,\vec k +\vec p -\vec q, \vec k', \vec k' -\vec p, \lambda}= \\
&& \quad \Big( \frac{\hat h_{\vec q}(t)}{2}+u_{\vec k + \vec p,\vec k +\vec p -\vec q,\lambda}\Big)
A_{\vec k, \vec k+\vec p; \vec k', \vec k' -\vec p}(\lambda, \Delta \lambda) \nonumber \\
&& \quad-  \Big( \frac{\hat h_{\vec q}(t)}{2}+u_{\vec k,\vec k -\vec q, \lambda} \Big)
A_{\vec k -\vec q, \vec k+\vec p-\vec q; \vec k', \vec k' -\vec p}(\lambda, \Delta \lambda) \,.\nonumber 
\end{eqnarray}
 The second contribution to the above commutator 
 arises from the term in $\mathcal H_{h,\lambda}(t)$ linear in $v$
  \begin{eqnarray}
 \label{B4}
&&   [X^\alpha_{\lambda, \Delta \lambda}(t), \hat{\mathcal H}_{h,\lambda}(t)]\big|_v 
=-   \frac{1}{N} \sum_{\vec k \vec k' \vec p \sigma} \Big\{
v_{\vec k, \vec k +\vec p -\vec q; \vec k', \vec k'- \vec p,\lambda} \nonumber
\\
&& \quad \times
[X^\alpha_{\lambda, \Delta \lambda}(t), \frac{\sigma}{2} 
:c^\dag_{\vec k \sigma} c^{}_{\vec k +\vec p -\vec q,\sigma}  c^\dag_{\vec k', -\sigma} 
c^{}_{\vec k' -\vec p,-\sigma}: ] 
  \nonumber  \\
&&  \quad  + \textrm{H.c.} \Big\} \, .
\end{eqnarray} 
Here, the evaluation has to be followed up by 
a decomposition into one-particle and 
two-particle contributions. This leads to renormalization contributions to $u_{\vec k, \vec k -\vec q, \lambda}$ and 
$v_{\vec k,\vec k +\vec p -\vec q, \vec k', \vec k' -\vec p, \lambda}$. However,
they turn out to be small and can be neglected. They are at least by a factor of order 
$O(U/ \Delta \varepsilon)$ smaller than the
contributions (\ref{B2}), (\ref{B3}), where $\Delta \varepsilon$ denotes an energy difference
of the order of the conduction electron band width.

 \subsection{Commutator  $[X_{\lambda, \Delta \lambda}(t), {\mathcal H}_{f,\lambda}]$ }
\label{appB.2} 
Due to  decompositions (\ref{33})  and (\ref{31}) of $X_{\lambda, \Delta \lambda}$ and $\mathcal H_{f,\lambda}$ 
one has to evaluate three different contributions to order $h(t)$.

\subsubsection{Commutator $ [X^\alpha_{\lambda, \Delta \lambda}(t), 
{\mathcal H}^\alpha_{f,\lambda}]$}
\label{appB.2.1}
The first commutator leads to renormalization contributions of order $U^2$: 
 \begin{eqnarray}
\label{B5}
&&[X^\alpha_{\lambda, \Delta \lambda}, {\mathcal H}^\alpha_{f,\lambda}] =
\frac{U}{2N^2} \sum_{\vec k \vec k' \vec k'', \vec p \vec p' \sigma}   
\Big\{
F_{\vec k \vec k' \vec k'', \vec p \vec p'}(\lambda, \Delta \lambda) \nonumber \\
&& \times 
c^\dag_{\vec k \sigma} c^{}_{\vec k + \vec p +\vec p', \sigma} \, :c^\dag_{\vec k' , -\sigma} c^{}_{\vec k' -\vec p',
-\sigma}: \,
: c^\dag_{\vec k'',-\sigma} c^{}_{\vec k'' -\vec p, -\sigma}: \nonumber \\
&& + \textrm{H.c.}   \Big\} \, ,
\end{eqnarray}  
 where 
 \begin{eqnarray}
\label{B6}
&& F_{\vec k \vec k' \vec k'' \vec p \vec p'}(\lambda, \Delta \lambda) =  \\
&& \quad
A_{\vec k, \vec k+ \vec p', \vec k', \vec k' -\vec p'}(\lambda, \Delta \lambda) \, 
\Theta_{\vec k+ \vec p' , \vec k + \vec p + \vec p', \vec k'', \vec k'' -\vec p, \lambda} \nonumber \\
&& \quad -
A_{\vec k +\vec p, \vec k+ \vec p +\vec p', \vec k', \vec k' -\vec p'}(\lambda, \Delta \lambda) \, 
\Theta_{\vec k, \vec k +\vec p , \vec k'', \vec k'' -\vec p, \lambda} \, .
\nonumber
\end{eqnarray} 
As before, the operator structure in (\ref{B5}) has to be reduced  to operators which  
appear in ${\mathcal H}_\lambda(t)$.  
Besides a term proportional to $c^\dag c$, also contributions with four creation and annihilation operators 
can be extracted from Eq.~(\ref{B5}).   The first contribution reads
\begin{eqnarray}
\label{B7}
&& 
[X^\alpha_{\lambda, \Delta \lambda},  {\mathcal H}^\alpha_{f,\lambda}] =
\frac{U}{2N^2} \sum_{\vec k \vec k' \vec k'' \vec p \vec p' \sigma}  
F_{\vec k \vec k' \vec k''; \vec p \vec p'}(\lambda, \Delta \lambda) \nonumber \\ 
&& 
\times\Big\{ \langle c^\dag_{\vec k \sigma} c^{}_{\vec k + \vec p +\vec p', \sigma}\rangle  \, 
\langle c_{\vec k' -\vec p', -\sigma}    c^{\dag}_{\vec k'', -\sigma} \rangle \,                 
:c^{\dag}_{\vec k' ,-\sigma} c^{}_{\vec k'' -\vec p,-\sigma}: \nonumber \\
&& + \langle c^\dag_{\vec k \sigma} c^{}_{\vec k + \vec p +\vec p', \sigma}\rangle  \, 
 \langle c^{\dag}_{\vec k',-\sigma} c^{}_{\vec k''- \vec p,-\sigma} \rangle 
 :c^{}_{\vec k' -\vec p', -\sigma}    c^{\dag}_{\vec k'' , -\sigma}:   \nonumber \\
&&+ \langle c^\dag_{\vec k', -\sigma}    c^{}_{\vec k'' -\vec p, -\sigma} \rangle \,                 
\langle c^{}_{\vec k'- \vec p',-\sigma} c^\dag_{\vec k'',-\sigma} \rangle \, 
:c^\dag_{\vec k \sigma} c^{}_{\vec k + \vec p +\vec p', \sigma}:   
   \nonumber \\
&& + \textrm{H.c.}   \Big\} \, ,
\end{eqnarray}  
where there are two options for the expectation values according to relation (\ref{10}):
   \begin{equation}
   \label{B8}
 \langle c_{\vec k \sigma}^\dag\, c^{}_{\vec k - {\vec p}, \sigma} \rangle =
\frac{\delta_{{\vec p},0}}{2} \langle n_{\vec k} \rangle 
+ \sigma  \delta_{{\vec p}, \pm \vec q}   \langle s^z_{\vec k, \pm \vec q}\rangle
  \, ,
 \end{equation}
 with either ${\vec p} =0$ or ${\vec p}=  \pm \vec q$. 
 For the choice $\vec p=0 $ the commutator (\ref{B7}) leads to a renormalization of  
the electronic one-particle energy $\varepsilon_{\vec k, \lambda}$, 
  \begin{eqnarray}
  \label{B9}
&& 
[X^\alpha_{\lambda, \Delta \lambda},  {\mathcal H}^\alpha_{f,\lambda}]\big|_{(p=0)} =
\sum_{\vec k,\sigma} \delta \varepsilon^{(1)}_{\vec k,\lambda } \, 
c_{\vec k \sigma}^\dag c^{}_{\vec k \sigma}  \,  ,
\end{eqnarray}
where $\delta \varepsilon^{(1)}_{\vec k,\lambda}$ is 
given by 
  \begin{eqnarray}
 \label{B10}
&& \delta \varepsilon^{(1)}_{\vec k,\lambda} = \\
&& \quad  \frac{U}{4N^2} \sum_{\vec k'   \vec p }  \Big[
F_{\vec k', \vec k, \vec k + \vec p; \vec p, -\vec p}(\lambda, \Delta \lambda) \, 
\langle n_{\vec k'}\rangle  \, (2- \langle n_{\vec k +\vec p}   \rangle)  \nonumber \\
&& \quad + 
F_{\vec k, \vec k', \vec k' +\vec p; \vec p, -\vec p}(\lambda, \Delta \lambda) \, 
\langle   n_{\vec k'}   \rangle \,                 
(2- \langle n_{\vec k' + \vec p} \rangle) \nonumber \\
&& \quad -
F_{\vec k', \vec k -\vec p, \vec k; \vec p, - \vec p}(\lambda, \Delta \lambda)  \,       
\langle n_{\vec k'}\rangle  \langle  n_{\vec k -\vec p}   \rangle 
 \Big] \nonumber
\end{eqnarray}
with $\langle c^\dag_{\vec k \sigma} c^{}_{\vec k \sigma}\rangle =\langle n_{\vec k} \rangle/2$.  
The second choice $\vec p = \pm \vec q$ leads to a renormalization 
of the effective field. The contribution to $u_{\vec k, \vec k -\vec q}(t)$ is 
\begin{eqnarray}
\label{B11}
[X^\alpha_{\lambda, \Delta \lambda},  {\mathcal H}^\alpha_{f,\lambda}]_{(p=\pm q)} &=& - \sum_{\vec k \sigma}
\Big\{ \delta u_{\vec k, \vec k -\vec q,\lambda}^{(2)}(t) 
 \frac{\sigma}{2} \, c^\dag_{\vec k \sigma} c^{}_{\vec k -\vec q, \sigma} \nonumber \\
 &&\qquad\quad+ \textrm{H.c.} \Big\} 
\end{eqnarray} 
 with
  \begin{eqnarray}
\label{B12}
&& \delta  u_{\vec k, \vec k -\vec q,\lambda}^{(2)}(t) =  - \frac{U}{2N^2} \sum_{\vec k' \vec p} 
 B^{(2)}_{\vec p \vec k'; \vec k \vec q}(\lambda, \Delta \lambda) \langle s^z_{\vec k', -\vec q}\rangle(t) \, ,
 \nonumber \\ 
 && 
\end{eqnarray}
where the pre-factor $ B^{(2)}_{\vec p \vec k';  \vec k \vec q}(\lambda, \Delta \lambda)$  is of order $U/\Delta \varepsilon$:
\begin{eqnarray}
\label{B13}
&&B^{(2)}_{\vec p \vec k' ; \vec k \vec q}(\lambda, \Delta \lambda)= \\
&& - F_{\vec p, \vec k, \vec k'; \vec k' -\vec k + \vec q, -(\vec k' -\vec k + \vec q)}(\lambda, \Delta \lambda)
\langle n_{\vec p} \rangle  \nonumber \\
&& - F_{\vec k', \vec k, \vec k + \vec p -\vec q; \vec p, -\vec p +\vec q}(\lambda, \Delta \lambda)
\big( 2- \langle n_{\vec k+\vec p -\vec q} \rangle \big) \nonumber \\
&& - F_{\vec p, \vec k', \vec k; \vec k -\vec k' - \vec q, -(\vec k -\vec k' - \vec q)}(\lambda, \Delta \lambda)
\langle n_{\vec p} \rangle  \nonumber \\
&&
+F_{\vec k', \vec k -\vec p, \vec k; \vec p, -\vec p +\vec q}(\lambda, \Delta \lambda)
\langle n_{\vec k-\vec p} \rangle  \nonumber \\
&&
+ F_{\vec k, \vec k' -\vec p, \vec k'; \vec p, -\vec p - \vec q}(\lambda, \Delta \lambda)
\langle n_{\vec k' -\vec p} \rangle  \nonumber \\
&&
- F_{\vec k, \vec k', \vec k + \vec p +\vec q; \vec p, -\vec p -\vec q}(\lambda, \Delta \lambda)
\big(2- \langle n_{\vec k' + \vec p + \vec q} \rangle \big)\,.  \nonumber 
\end{eqnarray}
Note that a common factor $\langle s^z_{\vec k', -\vec q}\rangle(t)$ was 
already extracted in Eq.~(\ref{B12}).  \\

The remaining contribution to commutator (\ref{B5}) with four creation and annihilation operators 
leads to  a renormalization of $\mathcal H_{f,\lambda}$ of order  $U^2/ \Delta\varepsilon$. Such contributions
have been left out from the very beginning. 

\subsubsection{Commutator $[X^\beta_{\lambda, \Delta \lambda}(t), 
{\mathcal H}^\alpha_{f,\lambda}]$}
\label{appB.2.2}

The evaluation of this commutator leads to renormalizations of 
$u_{\vec k, \vec k -\vec q, \lambda}$ and $v_{\vec k, \vec k +\vec p -\vec q; \vec k', \vec k'-\vec p,\lambda}$: 
\begin{eqnarray}
\label{B14}
&&[X^\beta_{\lambda, \Delta \lambda}(t), 
{\mathcal H}^\alpha_{f,\lambda}] = -  \sum_{\vec k \sigma}  \Big(\delta u^{(3)}_{\vec k, \vec k -\vec q, \lambda}
 \frac{\sigma}{2}\,  
c^\dag_{\vec k,\sigma} c^{}_{\vec k -\vec q, \sigma} + \mathrm{H.c.} \Big) \nonumber \\
&& \quad \qquad -\frac{1}{N} \sum_{\vec k \vec k'\vec p \sigma} \Big[
\delta v^{(2)}_{\vec k, \vec k +\vec p -\vec q; \vec k', \vec k'-\vec p,\lambda}   \nonumber\\
&& \quad \qquad\times \frac{\sigma}{2}
: c^\dag_{\vec k\sigma} c^{}_{\vec k + \vec p -\vec q, \sigma}\, c^\dag_{\vec k',-\sigma} c^{}_{\vec k' -\vec p,-\sigma}:
+ \mathrm{H.c.}
\Big]
\end{eqnarray}
with 
\begin{eqnarray}
\label{B15} 
&& \delta u^{(3)}_{\vec k, \vec k -\vec q, \lambda} = - \frac{2U}{N}
\sum_{\vec k'}  \Theta_{\vec k', \vec k' + \vec q; \vec k, \vec k-\vec q, \lambda}
B_{\vec k' + \vec q, \vec k'}(\lambda, \Delta \lambda)    \nonumber \\
&&\qquad \times \big(
\langle n_{\vec k' + \vec q, -\sigma}\rangle - \langle n_{\vec k', -\sigma}\rangle\big) \nonumber \\
\label{B16}
&& \delta v^{(2)}_{\vec k, \vec k +\vec p -\vec q; \vec k', \vec k'-\vec p,\lambda} = \nonumber \\
&& \qquad2U \Big( \Theta_{\vec k -\vec q, \vec k+ \vec p -\vec q; \vec k', \vec k' -\vec p,\lambda} 
B_{\vec k, \vec k- \vec q}(\lambda, \Delta \lambda) \nonumber \\
&& \qquad- 
\Theta_{\vec k, \vec k+\vec p ; \vec k', \vec k' -\vec p,\lambda} 
B_{\vec k +\vec p, \vec k+ \vec p - \vec q}(\lambda, \Delta \lambda) 
\Big)
\, , 
\end{eqnarray}
where 
$n_{\vec k \sigma} = \langle  c^\dag_{\vec k, \sigma} c^{}_{\vec k \sigma}  \rangle $. 
Inserting (\ref{36}) for $B_{\vec k, \vec k + \vec q}(\lambda, \Delta \lambda)$, we can 
again extract a common factor
 $\langle s^z_{\bar{\vec p}, - \vec q}\rangle $ from equations (\ref{B15}), (\ref{B16}). We find
\begin{eqnarray}
\label{B17}
&&  \delta u^{(3)}_{\vec k, \vec k -\vec q, \lambda} = - \frac{2U}{N^2}
\sum_{\vec p, \bar{\vec p}}  B^{(3)}_{\vec p \bar{\vec p}; \vec k \vec q}
(\lambda, \Delta \lambda) \langle s^z_{\bar{\vec p}, -\vec q}\rangle \,, \\\label{xyz1}
&& \delta v^{(2)}_{\vec k, \vec k +\vec p -\vec q; \vec k', \vec k'-\vec p, \lambda} =
\frac{2U}{N} \sum_{\bar{\vec p}} D^{(2)}_{\bar{\vec p} ; \vec k \vec k' \vec p\vec q}(\lambda, \Delta \lambda)
\langle s^z_{\bar{\vec p}, -\vec q}\rangle\nonumber\\&&
 \end{eqnarray}
 with 
 \begin{eqnarray}
&& B^{(3)}_{\vec p \bar{\vec p}; \vec k \vec q}(\lambda, \Delta \lambda) =
  \Theta_{\vec p, \vec p + \vec q; \vec k, \vec k- \vec q} 
\big(\langle n_{\vec p + \vec q, -\sigma}\rangle -\langle n_{\vec p, -\sigma} \rangle\big) \nonumber \\
&& \qquad\times
\big( A_{\vec p + \vec q; \vec p; \bar{\vec p}, \bar{\vec p} +\vec q}(\lambda,\Delta \lambda) -
A_{\vec p + \vec q, \vec p}(\lambda, \Delta \lambda)\big)
 \end{eqnarray}
 and
 \begin{eqnarray}
 && D^{(2)}_{\bar{\vec p} ; \vec k \vec k' \vec p\vec q}(\lambda, \Delta \lambda) =
 \Theta_{\vec k- \vec q, \vec k +\vec p -\vec q, \vec k', \vec k'-\vec p,\lambda} \\
 && \qquad \times 
 \Big( A_{\vec k, \vec k- \vec q; \bar{\vec p}, \bar{\vec p} + \vec q}(\lambda, \Delta \lambda)
 - A_{\vec k, \vec k -\vec q}(\lambda, \Delta \lambda) \Big) \nonumber \\
 && \qquad - 
  \Theta_{\vec k, \vec k +\vec p, \vec k', \vec k'-\vec p,\lambda} \nonumber\\
 && \qquad \times 
 \Big( A_{\vec k +\vec p, \vec k +\vec p - \vec q; \bar{\vec p}, \bar{\vec p} + \vec q}(\lambda, \Delta \lambda)
 - A_{\vec k+ \vec p, \vec k +\vec p  -\vec q}(\lambda, \Delta \lambda) \Big)\,.\nonumber
 \end{eqnarray}

\subsubsection{Commutator $[X^\alpha_{\lambda, \Delta \lambda}(t), 
{\mathcal H}^\beta_{f,\lambda}] $}
\label{appB.2.3}

In analogy to the last commutator we obtain the  renormalization 
contributions 
\begin{eqnarray}
\label{B18}
&& \delta u^{(4)}_{\vec k, \vec k -\vec q, \lambda} =  \frac{2U}{N}
\sum_{\vec k'}  A_{\vec k' -\vec q, \vec k'; \vec k, \vec k-\vec q} (\lambda, \Delta \lambda) \, 
\varphi_{\vec k', \vec k' -\vec q, \lambda}    \nonumber \\
&&\qquad \times \big(
\langle n_{\vec k', -\sigma} \rangle - \langle n_{\vec k' -\vec q, -\sigma}\rangle\big)  
\end{eqnarray}
and 
\begin{eqnarray}
\label{B19}
&& \delta v^{(3)}_{\vec k, \vec k +\vec p -\vec q; \vec k', \vec k'-\vec p,\lambda} =  \\
&& \quad-2U \Big( A_{\vec k -\vec q, \vec k+ \vec p -\vec q; \vec k', \vec k' -\vec p}(\lambda, \Delta \lambda) \, 
\varphi_{\vec k, \vec k- \vec q, \lambda} \nonumber \\
&& \quad- 
A_{\vec k, \vec k+\vec p ; \vec k', \vec k' -\vec p}(\lambda,\Delta \lambda) \, 
\varphi_{\vec k +\vec p, \vec k+ \vec p - \vec q,\lambda}
\Big)
\, , \nonumber
\end{eqnarray}
where the quantity $\varphi_{\vec k, \vec k -\vec q, \lambda}= 
 \varphi_{\vec k, \vec k -\vec q, \lambda}(t)$ was already defined in Eq.~(\ref{32}).
Due to this relation we can extract a common factor $\langle s^z_{\bar{\vec p}, -\vec q}\rangle $ in 
Eqs.~(\ref{B18}) and (\ref{B19}), and obtain:
\begin{eqnarray}
\label{B20}
&&  \delta u^{(4)}_{\vec k, \vec k -\vec q, \lambda} =  \frac{2U}{N^2}
\sum_{\vec p, \bar{\vec p}}   B^{(4)}_{\vec p \bar{\vec p}; \vec k \vec q}(\lambda, \Delta \lambda) 
 \langle s^z_{\bar{\vec p}, -\vec q}\rangle\,, \\\label{xyz2}
 && \delta v^{(3)}_{\vec k, \vec k +\vec p -\vec q; \vec k', \vec k'-\vec p,\lambda}  =
 -\frac{2U}{N} \sum_{\bar{\vec p}}  D^{(3)}_{\bar{\vec p}; \vec k \vec k' \vec p \vec q}
 \langle s^z_{\bar{\vec p}, -\vec q}\rangle \,,\qquad 
\end{eqnarray}
where
\begin{eqnarray}
&& B^{(4)}_{\vec p \bar{\vec p}; \vec k \vec q}(\lambda, \Delta \lambda) =
A_{\vec p -\vec q, \vec p; \vec k, \vec k-\vec q} (\lambda, \Delta \lambda) \\
&& \times
\big( \Theta_{\vec p, \vec p- \vec q; \bar{\vec p}, \bar{\vec p}+ \vec q, \lambda} -
 \Theta_{\vec p, \vec p- \vec q, \lambda}
\big)\big(\langle n_{\vec p, -\sigma}\rangle -\langle n_{\vec p -\vec q, -\sigma}\rangle\big)\,, \nonumber \\
&&  D^{(3)}_{\bar{\vec p}; \vec k \vec k' \vec p \vec q} = 
A_{\vec k -\vec q, \vec k+ \vec p-\vec q; \vec k', \vec k'-\vec p}(\lambda, \Delta \lambda) \nonumber \\
&& \quad \times 
\big( \Theta_{\vec k, \vec k- \vec q, \bar{\vec p}, \bar{\vec p}+ \vec q,\lambda}
- \Theta_{\vec k, \vec k- \vec q,\lambda}
\big) \nonumber \\
&& \quad  -
A_{\vec k, \vec k+ \vec p; \vec k', \vec k'-\vec p}(\lambda, \Delta \lambda) \nonumber \\
&& \quad \times 
\big( \Theta_{\vec k + \vec p, \vec k+ \vec p - \vec q, \bar{\vec p}, \bar{\vec p}+ \vec q,\lambda}
- \Theta_{\vec k + \vec p, \vec k+ \vec p - \vec q,\lambda}
\big) \nonumber  \, .\nonumber
\end{eqnarray}


\subsection{Commutator $  [X_{\lambda, \Delta \lambda}, [ X_{\lambda, \Delta \lambda}, 
 {\mathcal H}_{0,\lambda}] ]$}  
  \label{B.3}
  
  First, by use of Eq.~(\ref{23}) this commutator  can be transformed to $- [X_{\lambda, \Delta \lambda}, 
   \mathbf Q_{\lambda -\Delta \lambda} {\mathcal H}_{f,\lambda} ]$, where 
   $X_{\lambda, \Delta \lambda}$ and ${\mathcal H}_{f,\lambda}$ consist of two parts.  
   For the commutator part with $X^\alpha_{\lambda, \Delta \lambda}$  and $\mathcal H^\alpha_{f,\lambda}$ 
   one starts from expression (\ref{34}) for $X^\alpha_{\lambda, \Delta \lambda}$ and 
    \begin{eqnarray}
   \label{B21}
&&   \mathbf Q_{\lambda -\Delta \lambda}{\mathcal H}^\alpha_{f, \lambda} = 
\frac{U}{2N} \sum_{\vec k \vec k' \vec p \sigma}  \Theta_{\vec k, \vec k+\vec p; \vec k', \vec k' -\vec p, \lambda}    
  \nonumber  \qquad\\
  && \qquad \times
\big( 1 - \Theta_{\vec k, \vec k+\vec p; \vec k', \vec k' -\vec p, \lambda - \Delta \lambda} 
 \big)
  \nonumber \\
&& \qquad \times  :c^\dag_{\vec k \sigma} \,c^{}_{\vec k + \vec p, \sigma} \, c^\dag_{\vec k', -\sigma} \,c^{}_{\vec k' - \vec p, -\sigma}:  \, . 
\end{eqnarray}
Note that both $X^\alpha_{\lambda, \Delta \lambda}$ 
and $ \mathbf Q_{\lambda -\Delta \lambda}{\mathcal H}^\alpha_{f, \lambda}$ contain a product of two 
$\Theta$ functions which restrict the allowed excitations to the small
energy shell between $\lambda$ and $\lambda -\Delta \lambda$.
Therefore only those contributions  to the  commutator are important, for which identical  excitation energies 
enter the two $\Theta$ function products. We get 
\begin{eqnarray}
\label{B22}
&& -[X^\alpha_{\lambda, \Delta \lambda}, \mathbf Q_{\lambda- \Delta \lambda}\mathcal H^\alpha_{f,\lambda}]] = \\
&& \qquad -
\frac{U}{2N^2} 
\sum_{\vec k \vec k' \vec p, \sigma}
\Big\{
G_{\vec k, \vec k', \vec k'+ \vec p, \vec p, -\vec p}(\lambda, \Delta \lambda) \nonumber \\
&&\qquad \times
c^\dag_{\vec k\sigma} c^{}_{\vec k\sigma}\, :c^\dag_{\vec k',-\sigma} c^{}_{\vec k'+\vec p, -\sigma}:
\, :c^\dag_{\vec k' + \vec p, -\sigma} c^{}_{\vec k', -\sigma}:  \nonumber \\
&&  \qquad+ \textrm{H.c.} \Big\}   \, , \nonumber
\end{eqnarray}
where we have introduced
\begin{eqnarray}
\label{B23}
&& G_{\vec k, \vec k', \vec k' + \vec p, \vec p, -\vec p}(\lambda, \Delta \lambda) = \\
&&  \qquad
A_{\vec k, \vec k- \vec p; \vec k', \vec k'+ \vec p}(\lambda, \Delta \lambda)   - 
A_{\vec k+ \vec p, \vec k; \vec k', \vec k'+ \vec p}(\lambda, \Delta \lambda) \, .
 \nonumber 
\nonumber \\
&& \nonumber
\end{eqnarray}
After reducing (\ref{B22}) to the one-particle part, we are led to the following renormalization of $ \varepsilon_{\vec k,\lambda }$:
  \begin{eqnarray}
   \label{B24}
 && \delta   \varepsilon^{(2)}_{\vec k,\lambda }  =   \\
 &&  \quad
   \frac{U}{4N^2} \sum_{\vec k'   \vec p' }  \Big[
 G_{\vec k', \vec k, \vec k+ \vec p; \vec p , -\vec p}(\lambda, \Delta \lambda) \langle n_{\vec k'}\rangle
\big(2- \langle n_{\vec k +\vec p}\rangle\big)  
\nonumber \\
&&\quad  +  G_{\vec k, \vec k' , \vec k' + \vec p; \vec p  , -\vec p}(\lambda, \Delta \lambda) 
\langle n_{\vec k'}\rangle \big (2- \langle n_{\vec k' +\vec p}\rangle \big)
\nonumber \\
&&  \quad- G_{\vec k', \vec k -\vec p, \vec k; \vec p , -\vec p}(\lambda, \Delta \lambda) \langle n_{\vec k'}\rangle
\langle n_{\vec k -\vec p}\rangle  \nonumber 
\Big]\,.  
\end{eqnarray}
Possible contributions to $u_{\vec k, \vec k - \vec q, \lambda}$ and 
$v_{\vec k, \vec k +\vec p -\vec q; \vec k', \vec k'-\vec p,\lambda}$ cancel. 
The commutator part of (\ref{B22}) with four fermion operators 
leads to higher order contributions to
$\mathcal H_{f,\lambda}$, which have been neglected. The remaining 'mixed' contributions 
$(\alpha \beta)$, $(\beta \alpha)$  to 
the commutator 
$[X_{\lambda, \Delta \lambda}, \mathbf Q_{\lambda- \Delta \lambda}\mathcal H_{f,\lambda}]]$, which are
of order $h(t)$,  are negligible  as well, since the energies of the two $\Theta$-function products do not coincide.  

Finally we consider the last commutator  in transformation (\ref{24}), $[ X_{\lambda, \Delta \lambda}, 
  [ X_{\lambda, \Delta \lambda},  \hat {\mathcal H}_{h,\lambda}]  ] $.  Here
   only the four fermion part $X^\alpha_{\lambda, \Delta \lambda}$ contributes in first order in $h(t)$. 
   Renormalization contributions to $u_{\vec k, \vec k -\vec q, \lambda}$ and 
   $v_{\vec k, \vec k +\vec p -\vec q; \vec k', \vec k'-\vec p, \lambda}$ are again expected to be small 
   due to the appearance of the two $\Theta$-function products from the two generators 
   $X^\alpha_{\lambda, \Delta \lambda}$.
  

\section{Renormalization contributions of Eqs.~(\ref{66}) and (\ref{67}) }
\label{appC}

In Sect.~V.D  the general result (\ref{61}) for $\langle s^z_{-\vec q}\rangle(t)$ was simplified by tracing back the 
$\vec k$-resolved expectation values $\langle s^z_{\vec k, -\vec q}\rangle(t)$
to the compact variable $\langle s^z_{-\vec q}\rangle(t)$. The resulting  time-independent
renormalization equations are given in Eqs.~(\ref{66}), (\ref{67}). The renormalization contributions on the 
right-hand sides read
  \begin{eqnarray}
 \label{C1}
&&   \delta u_{\vec k,\vec k -\vec q, \lambda}^{0(1)}=  \\
&& \qquad  - \frac{1}{N} \sum_{\vec k'} \Big( U + \frac{1}{\chi(\vec q, \omega)}
 + u^0_{\vec k', \vec k' -\vec q, \lambda} \Big) \nonumber \\
&& \qquad \times A_{\vec k, \vec k-\vec q; \vec k' -\vec q, \vec k'}(\lambda, \Delta \lambda) \nonumber \\
 && \qquad \times \big(
 \langle c_{\vec k',-\sigma}^\dag c_{\vec k', -\sigma}\rangle
-   \langle c_{\vec k' -\vec q,-\sigma}^\dag c_{\vec k' -\vec q, -\sigma}\rangle
\big) \,,\nonumber \\
\label{C2}
 && \sum_{n=2}^4 \delta  u_{\vec k, \vec k -\vec q,\lambda}^{0(n)} =  \\
 && \qquad  - \frac{U}{2N^2} \sum_{\vec p \bar{\vec p}} 
 \big( B^{(2)}_{\vec p \bar{\vec p}; \vec k \vec q}(\lambda, \Delta \lambda) 
 + 4 B^{(3)}_{\vec p \bar{\vec p}; \vec k \vec q}(\lambda, \Delta \lambda) \nonumber \\
&& \qquad  -4 B^{(4)}_{\vec p \bar{\vec p}; \vec k \vec q}(\lambda, \Delta \lambda) 
 \big) \nonumber 
 \end{eqnarray}
and
  \begin{eqnarray}
 \label{C3}
&&   \delta v^{0(1)}_{\vec k,\vec k +\vec p -\vec q, \vec k', \vec k' -\vec p, \lambda}= \\
&&  
\Big( U+ \frac{1}{\chi(\vec q, \omega)} + u^0_{\vec k + \vec p,\vec k +\vec p -\vec q,\lambda} \Big)
A_{\vec k, \vec k+\vec p; \vec k', \vec k' -\vec p}(\lambda, \Delta \lambda) \nonumber \\
&& - \Big(  U+ \frac{1}{\chi(\vec q, \omega)}    + u^0_{\vec k,\vec k -\vec q, \lambda} \Big) 
A_{\vec k -\vec q, \vec k+\vec p-\vec q; \vec k', \vec k' -\vec p}(\lambda, \Delta \lambda) \, ,
\nonumber
\end{eqnarray}
 \begin{eqnarray}
 \label{C4}
&& \sum_{n=2,3} \delta v^{0(n)}_{\vec k,\vec k +\vec p -\vec q, \vec k', \vec k' -\vec p, \lambda} =  \\
&& \quad\frac{2U}{N} \sum_{\bar{\vec p}}\Big( D^{(2)}_{\bar{\vec p}; \vec k \vec k' \vec p \vec q}(\lambda, \Delta \lambda)
- D^{(3)}_{\bar{\vec p}; \vec k \vec k' \vec p \vec q}(\lambda, \Delta \lambda)
\Big) \nonumber\,,
\end{eqnarray}
where
\begin{equation}
\label{C5}
\frac{\hat h_{\vec q}(t)}{2} = \Big(U + \frac{1}{\chi(\vec q, \omega)}\Big) \frac{\langle s^z_{-\vec q}\rangle(t)}{N} \,.
\end{equation}  


\end{appendix}



\bibliographystyle{unsrt}

\begin{thebibliography}{10}

\bibitem{BSW89}
N.~E. Bickers, D.~J. Scalapino, and S.~R. White.
\newblock {\em Phys. Rev. Lett.}, 62:961, 1989.

\bibitem{BK61}
G.~Baym and L.~P. Kadanoff.
\newblock {\em Phys. Rev.}, 124:287, 1961.

\bibitem{KB62}
L.~P. Kadanoff and G.~Baym.
\newblock {\em Quantum Statistical Mechanics}.
\newblock Benjamin/Cumming Publishing Company, Reading, Massachusetts, 1962.

\bibitem{GV08}
G.~Giuliani and G.~Vignale.
\newblock {\em Quantum Theory of the Electron Liquid}.
\newblock Cambridge University Press, Cambridge UK, 2008.

\bibitem{VT97}
Y.~M. Vilk and A.~M.~S. Tremblay.
\newblock {\em J. Phys. I (France)}, 7:1309, 1997.

\bibitem{BSW94}
N.~Bulut, D.~J. Scalapino, and S.~R. White.
\newblock Technical report, University of Illinois, 1994.

\bibitem{VCT94}
Y.~M. Vilk, L.~Chen, and A.~M.~S. Tremblay.
\newblock {\em Phys. Rev. B}, 49:13267, 1994.

\bibitem{JSDB05}
M. Jema\"\i, P. Schuck, J. Dukelsky, and R. Bennaceur.
\newblock {\em Phys. Rev. B}, 71:085115, 2005.

\bibitem{KS94}
P.~Kr\"uger and P.~Schuck.
\newblock {\em Europhys. Lett.}, 27:395, 1994.

\bibitem{HMDS02}
J.~G. Hirsch, A.~Mariano, J. Dukelsky, and P. Schuck.
\newblock {\em Ann. Phys.}, 296:187, 2002.

\bibitem{BHS02}
K.~W. Becker, A.~H\"ubsch, and T.~Sommer.
\newblock {\em Phys. Rev. B}, 66:235115, 2002.

\bibitem{SHB09b}
S.~Sykora, A.~H\"ubsch, and K.~W. Becker.
\newblock {\em Europhys. Lett.}, 85:57003, 2009.

\bibitem{HB03}
A.~H\"ubsch and K.~W. Becker.
\newblock {\em Eur. Phys. J. B}, 33:1434, 2003.

\bibitem{SHBWF05}
S.~Sykora, A.~H{\"u}bsch, K.~W. Becker, G.~Wellein, and H.~Fehske.
\newblock Single-particle excitations and phonon softening in the
  one-dimensional spinless holstein model.
\newblock {\em Phys. Rev. B}, 71:045112, 2005.

\bibitem{SHB06}
S.~Sykora, A.~H{\"u}bsch, and K.~W. Becker.
\newblock {\em Europhys. Lett.}, 76:644, 2006.

\bibitem{PFB11}
V.-N. Phan, H.~Fehske, and K.~W. Becker.
\newblock {\em Europhys. Lett.}, 95:17006, 2011.

\bibitem{PBF13}
V.-N. Phan, K.~W. Becker, and H.~Fehske.
\newblock {\em Phys. Rev. B}, 88:205123, 2013.

\bibitem{PF12}
V.-N. Phan and H.~Fehske.
\newblock {\em New J. Phys.}, 14:075007, 2012.

\bibitem{PBF10}
V.-N. Phan, K.~W. Becker, and H.~Fehske.
\newblock {\em Phys. Rev. B}, 81:205117, 2010.

\bibitem{ESBF12}
S.~Ejima, S.~Sykora, K.~W. Becker, and H.~Fehske.
\newblock {\em Phys. Rev. B}, 86:155149, 2012.

\bibitem{PMB10}
V.-N. Phan, A.~Mai, and K.~W. Becker.
\newblock {\em Phys. Rev. B}, 82:045101, 2010.

\bibitem{SB13}
S.~Sykora and K.~W. Becker.
\newblock {\em Sci. Rep.}, 3:2691, 2013.

\bibitem{Gu63}
M.~C. Gutzwiller.
\newblock {\em Phys. Rev. Lett.}, 10:159, 1963.

\bibitem{Hu63}
J.~Hubbard.
\newblock {\em Proc. Roy. Soc. London, Ser. A}, 276:238, 1963.

\bibitem{Ka63}
J. Kanamori.
\newblock {\em Prog. Theor. Phys.}, 30:275, 1963.

\bibitem{Tak99}
M.~Takahashi.
\newblock {\em Thermodynamics of One-Dimensional Solv\-able Models}.
\newblock Cambridge University Press, Cambridge, 1999.

\bibitem{EFGKK05}
F.~H.~L. Essler, H. Frahm, F. G{\"o}hmann, A. Kl{\"u}mper, and
  V.~E. Korepin.
\newblock {\em The One-Dimensional Hubbard Model}.
\newblock Cambridge University Press, Cambridge, 2005.

\bibitem{We94}
F. Wegner. \newblock{\em Ann. Phys. (Leipzig)}, 3:77, 1994. 

\bibitem{Keh06}
S. Kehrein.
\newblock {\em The Flow Equation Approach to Many-Particle Systems}.
\newblock Springer Tracts in Modern Physics 217, Springer Berlin, 2006.

\bibitem{ZaDo11}
M. Zapalska and T. Doma\'nski.
\newblock{\em Phys. Rev. B}, 84:174520, 2011.  

\bibitem{FrKe10}
P. Fritsch and S. Kehrein. 
\newblock {\em Phys. Rev. B}, 81:035113, 2010.

\bibitem{KDU12}
H. Krull, N.~A. Drescher, and G.~S. Uhrig.
\newblock {\em Phys. Rev. B}, 86:125113, 2012.

\bibitem{VMM13}
A. Verdeny, A. Mielke, and F. Mintert.
\newblock {\em Phys. Rev. Lett.}, 111:175301, 2013. 


\bibitem{HSB08}
A. H\"ubsch, S. Sykora, and K.~W. Becker.
\newblock arXiv:0809.3360.

\bibitem{Pe66}
D.~Penn.
\newblock {\em Phys. Rev.}, 142:350, 1966.

\bibitem{Hi85a}
J.~E. Hirsch.
\newblock {\em Phys. Rev. B}, 31:4403, 1985.

\bibitem{BR70}
W.~F. Brinkman and T.~M. Rice.
\newblock {\em Phys. Rev. B}, 2:4302, 1970.

\bibitem{KR86}
G.~Kotliar and A.~E. Ruckenstein.
\newblock {\em Phys. Rev. Lett.}, 57:1362, 1986.

\bibitem{DFB92}
M.~Deeg, H.~Fehske, and H.~B\"uttner.
\newblock {\em Z. Phys. B}, 88:283, 1992.

\bibitem{MW66}
N.~D. Mermin and H.~Wagner.
\newblock {\em Phys. Rev. Lett.}, 17:1133, 1966.

\bibitem{Mah00}
G.~D. Mahan.
\newblock {\em Many-particle physics}.
\newblock Kluwer Academic/Plenum Publishers, New York, 2000.

\bibitem{DFKTI94}
M.~Deeg, H.~Fehske, S.~K\"orner, S.~Trimper, and D.~Ihle.
\newblock {\em Z. Phys. B}, 95:87, 1994.

\bibitem{ZIBF11}
B. Zenker, D. Ihle, F.~X. Bronold, and H. Fehske.
\newblock {\em Phys. Rev. B}, 83:235123, 2011.

\end{thebibliography}


\end{document}